\documentclass{aa}  

\usepackage{graphicx}
\usepackage{txfonts}
\usepackage{bm} 
\usepackage{chemformula} 
\let\ce\ch 
\usepackage{xspace} 
\newcommand{\kms}{\,km\,s$^{-1}$\xspace} 
\newcommand{\mm}{\,mm\xspace} 
\newcommand{\hrs}{\,hrs\xspace} 
\newcommand{\K}{\,K\xspace} 
\newcommand{\mK}{\,mK\xspace} 
\newcommand{\GHz}{\,GHz\xspace} 
\newcommand{\kHz}{\,kHz\xspace} 
\newcommand{\benzo}{c-\ce{C6H5CN}\xspace} %
\usepackage{booktabs} 
\usepackage{array} 
\usepackage{upgreek} 
\usepackage{orcidlink} 
\usepackage{utfsym} 
\usepackage{subfigure} 
\usepackage{multicol} 
\usepackage{subcaption} 
\usepackage{multirow} 
\begin{document}

   \title{Aromatic rings in the Central Molecular Zone: Benzonitrile}


    \titlerunning{Aromatic rings in the Central Molecular Zone}
    \authorrunning{V.M. Rivilla et al.}
    
   \author{V.M. Rivilla
          \inst{1},
    D. San Andrés 
          \inst{1,2},
    M. Sanz-Novo
          \inst{3,1},
    L. Colzi  
          \inst{1},
    I. Jiménez-Serra
          \inst{1},
A. López-Gallifa
\inst{1},
A. Martínez-Henares
\inst{1},
A. Megías
\inst{1},
S. Martín
\inst{4,5},
B. Tercero
\inst{6},
S. Zeng
\inst{7},
J. Loreau
\inst{8},
M. Ben Khalifa
\inst{8},
M. A. Requena-Torres
\inst{9,10},
   \and
P. de Vicente
\inst{11,12}
          }

   \institute{Centro de Astrobiolog{\'i}a (CAB), CSIC-INTA, Carretera de Ajalvir km 4, Torrej{\'o}n de Ardoz, 28850, Madrid, Spain\\
              \email{vrivilla@cab.inta-csic.es}
         \and
        Departamento de F{\'i}sica de la Tierra y Astrof{\'i}sica, Facultad de Ciencias F{\'i}sicas, Universidad Complutense de Madrid, 28040 Madrid, Spain
         \and
        Center for Astrochemical Studies, Max-Planck-Institut f{\"{u}}r extraterrestrische Physik, Giessenbachstrasse 1, Garching bei Munchen, 85748, Germany 
        \and
        European Southern Observatory, Alonso de C\'ordova 3107, Vitacura 763 0355, Santiago, Chile
         \and
        Joint ALMA Observatory, Alonso de C\'ordova 3107, Vitacura 763 0355, Santiago, Chile
         \and
        Observatorio Astron\'omico Nacional (OAN-IGN), Calle Alfonso XII, 3, 28014 Madrid, Spain   
         \and
        Star and Planet Formation Laboratory, Cluster for Pioneering Research, RIKEN, 2-1 Hirosawa, Wako, Saitama, 351-0198, Japan
        \and
        KU Leuven, Department of Chemistry, Celestijnenlaan 200F, B-3001 Leuven, Belgium
        \and
        University of Maryland, College Park, ND 20742-2421 (USA)
         \and
        Department of Physics, Astronomy and Geosciences, Towson University, Towson, MD 21252, USA  
        \and
        Observatorio de Yebes (OY-IGN), Cerro de la Palera SN, Yebes, Guadalajara, Spain 
}

   \date{Received ; accepted }

 
  \abstract
{In recent years, several aromatic molecules (i.e., rings based on benzene, c-\ce{C6H6}) have been reported towards the cold molecular cloud TMC-1, with benzonitrile (the CN-derivative of benzene, \benzo) being the one also identified in other nearby cold clouds with similar physical properties. The derived molecular abundances  differ significantly from those predicted by current chemical models, indicating that 
its chemistry is still poorly understood. This motivates the search for these compounds in other physically distinct sources beyond the solar vicinity.}
   {In this work, we present new detections of \benzo in two warmer molecular clouds of the Central Molecular Zone (CMZ) of our Galaxy, G+0.693-0.027 and G+0.633-0.0604.}
   {We used Yebes 40m ultra-deep surveys towards both CMZ clouds covering the 31$-$50 GHz range, and applied Local Thermodynamic Equilibrium (LTE) and non-LTE analysis to identify \benzo and derive the physical parameters of the emission.}
   {We derived column densities of $N$=(7.4$\pm$0.5)$\times$10$^{12}$ and (2.60$\pm$0.13)$\times$10$^{12}$ cm$^{-2}$, which translate into very similar molecular abundances relative to \ce{H2} of (6$\pm$1)$\times$10$^{-11}$ and (4.3$\pm$0.9)$\times$10$^{-11}$, respectively, consistent with those found towards cold clouds in the Galactic disk.
   We found that the ratio between the unsaturated carbon chain \ce{HC7N} and \benzo shows a dependence with the environmental conditions, being lower (2.15-2.4) in both CMZ clouds compared to colder sources (4.5$-$30). This  points to local effects on the formation/destruction of these species, and a preference of aromatic chemistry in the CMZ compared with linear carbon chains.}
   {This work confirms that \benzo is widespread in our Galaxy and can survive in harsher environments (high kinetic temperatures, strong shocks and enhanced cosmic-ray ionization rates) than those in Galactic cold clouds, which suggests that aromatics are stable and abundant species that can significantly contribute  to the total budget of interstellar carbon in molecular clouds in diverse environments. A top-down scenario, in which aromatic species are produced through the fragmentation and processing of carbonaceous solids or larger aromatics, agrees well with the strikingly nearly-constant behavior of their molecular abundances with molecular size.}
   \maketitle
%

\section{Introduction}

Carbon (C) is an essential chemical element for life because it offers unique capabilities to build up the macromolecules on which biochemistry is based, such as proteins, lipids, nucleic acids, and carbohydrates. C-based compounds provide an excellent balance between molecular diversity, stability and reactivity (\citealt{petkowski2020}). 
For this reason, to understand how life could have been triggered on Earth and how it might arise in other planetary systems, it is important to understand the reservoir and evolution of the C budget in the 
interstellar medium (ISM).
 
It is currently thought that a substantial fraction of interstellar C is present in the form of Polycyclic Aromatic Hydrocarbons (PAHs), which are organic compounds made up with benzene rings (c-\ce{C6H6})\footnote{Formally, the definition of an aromatic compound requires a cyclically conjugated molecular entity that fullfils the Hückel rule (4$n$+2 deslocalized $\pi$ electrons; e.g. $n$=1 for benzene; \citealt{iupac1997}).}.
Different estimates suggest that PAHs can contain between 10$\%$ and 25$\%$ of the interstellar C (\citealt{dwek1997,chiar2013}). 
As a consequence, the presence of these molecules has important implications for the physics and chemistry of the ISM, since the largest PAHs might play an important role in the growth of carbonaceous dust grains that can potentially transfer prebiotic material into new planetary systems \citep{Piacentino2025}.

PAHs have been detected towards a plethora of different interstellar environments, including galactic and extragalactic sources (e.g., \citealt{tielens2008,garcia-bernete2022,chown2024_orion,chown2025_galaxies}), through their emission bands at mid-infrared wavelengths produced by vibrational modes of
aromatic C$-$C and C$-$H bonds. However, this technique does not allow to identify specific PAHs, with the only exception of the six-membered ring benzene (c-\ce{C6H6}), which was detected in several circumstellar material around evolved stars through the observation of an absorption feature due to a bending mode near 14.85 $\upmu$m using the Infrared Space Observatory (ISO) and the Spitzer space telescope
(\citealt{cernicharo2001,kraemer2006,malek2012}); and in emission towards disks around very low-mass stars using the James Webb Space Telescope (e.g. \citealt{tabone2023,arabhavi2024}).

Only recently, two ultra-deep spectral surveys towards the cold dark cloud TMC-1 (QUIJOTE\footnote{Q-band Ultrasensitive Inspection Journey to the Obscure TMC-1
Environment.} and GOTHAM\footnote{GBT Observations of TMC-1: Hunting Aromatic Molecules.}), have successfully identified, for the first time, several specific aromatic species using rotational spectroscopy at cm/mm wavelengths.
First, \citet{mcguire2018} detected the CN-substitute of benzene, benzonitrile (\benzo). 
The CCH derivative of benzene (c-\ce{C6H5CCH}) and benzyne (o-\ce{C6H4}) have also been recently identified (\citealt{loru2023} and \citealt{cernicharo2021_benzyne}, respectively).
Moreover, several five-membered rings\footnote{We note that these five-carbon hydrocarbon cycles do not fulfill the criteria of aromaticity (see footnote 1), neither indene and derivatives, mentioned below, but we include them here for completeness.} were also detected: cyclopentadiene (c-\ce{C5H6}, \citealt{cernicharo2021_indene}); cyanocyclopentadiene (c-\ce{C5H5CN}; \citealt{mccarthy2021,lee2021}), ethynylcyclopentadiene (c-\ce{C5H5CCH}; \citealt{cernicharo2021_ethynylcyclopentadiene}), and fulvenallene (c-\ce{C5H4CCH2}; \citealt{cernicharo2022_fulvenallene}).

Remarkably, beyond single cycles, compounds with more than one ring have also been identified towards TMC-1. Regarding two rings, indene (c-\ce{C9H8}) and its CN-substitute were detected (\citealt{cernicharo2021_indene,burkhardt2021_indene,sita2022}), along with two isomers of cyanonaphtalenes (c-\ce{C10H7CN}; \citealt{mcguire2021,cernicharo2024}). Moreover,  the three-rings cyanoacenaphthylene (c-\ce{C12H7CN}) and 1$H$-cyclopent[$cd$]indene (c-\ce{C11H8}) have been reported by \citet{cernicharo2024} 
and \citet{fuentetaja2026}, respectively.
\citet{Wenzel2024a, Wenzel2024b} identified three isomers of cyanopyrene, with 4 rings (c-\ce{C16H9CN}; \citealt{Wenzel2024a, Wenzel2024b}), and also cyanocoronene, with 7 rings (c-\ce{C24H11CN}, \citealt{wenzel2025}). 

Beyond TMC-1, \benzo has been identified towards a handful of other cold clouds ($T_{\rm kin}\sim$10 K) in the Galactic disk within the Solar vicinity ($d<$ 500 pc)  \citep{burkhardt2021, agundez2023a}. 
All these detections are likely showing us the tip of the iceberg of the actual zoo of aromatic compounds present in interstellar molecular clouds.

One of the remaining open questions is how these PAHs (along with other carbon-cycles composing certain aromatic molecules) can be synthetised in space.
In this regard, two main paradigms have been proposed: i) bottom-up, in which larger PAHs might be formed {\it in situ} in molecular clouds by the combination of smaller species (e.g., \citealt{doddipatla2021,Kaiser&Hansen2021,cernicharo2022_fulvenallene,Reizer2022,cernicharo2023_maps,garciadelaConcepcion2023,yang2024,castineira2024,Mallo2025}); and ii) top-down, in which simple PAHs would be the result of destruction of dust carbonaceous grains by collisions or interstellar shocks  (\citealt{jones1996,scott1997,merino2014}), and/or fragmentation of large PAHs formed in the hot circumstellar medium (CSM) around evolved stars (e.g. \citealt{zhao2019}). In both cases, it is also  unclear whether the molecules form predominantly in the gas phase or on the surface of grains.

To discriminate between the proposed scenarios, it is essential to achieve detections of these simple aromatic species across a broad range of interstellar environments spanning diverse physical conditions (e.g., evolutionary stages, gas temperatures, turbulence levels, cosmic-ray ionisation rates, etc), since these parameters are expected to imprint distinct signatures on the resulting aromatic chemistry.
In fact, observations of sources beyond our Galaxy neighborhood is required to confirm whether these aromatic species are genuinely ubiquitous throughout the ISM.
In particular, observations within the more extreme conditions of the inner $\sim$100 pc of the Galaxy, known as the Central Molecular Zone (CMZ), will be especially informative about its survival within harsh  environments
The physical conditions of the CMZ molecular clouds differ substantially from those in cold clouds in the Galactic disk. First, they exhibit higher gas kinetic temperatures ranging from $\sim$50 to $\sim$150 K (\citealt{guesten1985,huettemeister1993,rodriguez-fernandez2001,ginsburg2016,krieger2017}).
Moreover, their chemistry can also be affected by the presence of large-scale shocks (e.g \citealt{martin-pintado2000,Requena-Torres2006,Martin2008}), and significantly enhanced cosmic-ray ionization rates (e.g. \citealt{goto2013,goto2014,yusef-zadeh2007,yusef-zadeh2013,yusef-zadeh2016}).

Recently, we have reported the first interstellar detection of a six-membered sulfur-bearing cyclic hydrocarbon,  2,5-cyclohexadien-1-thione (c-\ce{C6H6S}), towards the CMZ molecular cloud G+0.693-0.027 (\citealt{Araki2026}), demonstrating that carbon rings can also be present in these environments. In this work, we present the detection of \benzo, which is considered a proxy of aromatic chemistry, towards two molecular clouds within the Sgr B2 complex, namely G+0.693-0.027 and G+0.633-0.0604.

\section{Observations}\label{sec:observations}

\subsection{G+0.693-0.027}

We have used an unbiased and ultrasensitive spectral line survey carried out towards the GC molecular cloud G+0.693-0.027 (hereafter G+0.693) located in the Sgr B2 complex. To search for c-\ce{C6H5CN} in this source, we have employed $Q$-band observations at centimeter wavelengths (31.075-50.424 GHz), conducted with the Yebes 40m radiotelescope (Guadalajara, Spain; project 21A014, PI: Rivilla). We used the position switching mode, centered at $\alpha_\text{ICRS}$ = $\,$17$^{\rm h}$47$^{\rm m}$22$^{\rm s}$, $\delta_\text{ICRS}$ = $\,-$28$^{\circ}$21$^{\prime}$27$^{\prime\prime}$, with the off position shifted by $\Delta\alpha_\text{ICRS}$~=~$-885$$^{\prime\prime}$ and $\Delta\delta_\text{ICRS}$~=~$290$$^{\prime\prime}$; namely at $\alpha_\text{ICRS}(\text{OFF}) = 17^\text{h}46^\text{m}14.92^\text{s}$, $\delta_\text{ICRS}(\text{OFF}) = -28^\text{o}16'36.99''$. The half power beam width (HPBW) of the Yebes 40$\,$m telescope varies between $\sim$35$^{\prime\prime}$$-$55$^{\prime\prime}$ (at 50 and 31 GHz, respectively; \citealt{tercero2021}). The line intensity of the spectra is presented in units of antenna temperature ($T_{\mathrm{A}}^{\ast}$) since the molecular emission towards G+0.693 is extended beyond the telescope primary beams (\citealt{Brunken2010, Jones2012, Li2020, Santa-Maria2021, Colzi2024}). Further details on these observations, including resolution and final noise levels of the molecular line survey, are given in \citet{Rivilla2023} and \citet{Sanz-Novo2023}.

\subsection{G+0.633-0.0604}

We have used data from a newly conducted ultra-sensitive spectral line survey towards the molecular cloud G+0.633-0.0604 (hereafter G+0.633), which is located in the southern region of Sgr B2, at $\sim$250$^{\prime\prime}$ (i.e. $\sim$10.4 pc in projected distance) from G+0.693. 
Similarly to G+0.693, we performed the analysis for \benzo in this cloud using data obtained with the Yebes 40m telescope (Guadalajara, Spain; project 24A010, PI: San Andr{\'e}s) along 25 night-shifts from February$-$April 2024, for a total observing time of $\sim$90\hrs. 
We performed position switching observations centered at $\alpha_\text{ICRS}= 17^\text{h}47^\text{m}21.18^\text{s}$, $\delta_\text{ICRS} = -28^\text{o}25'36.99''$. The same Off position than in G+0.693 was used, which translates into a shift with respect to G+0.633 coordinates of $\Delta\alpha_\text{ICRS} = -874''$ and $\Delta\delta_\text{ICRS} = 540''$.
These observations cover the spectral range 31.07$-$50.42\GHz, with the half power beam width (HPBW) varying from $\sim$35$''$ to $\sim$55$''$.
The spectra were smoothed into a resolution of $\sim$256\kHz ($\sim$1.54$-$2.56\kms in the observed range), which is enough to properly resolve the observed linewidths of $\sim$10\kms. The root mean square (rms) noise levels achieved are between 0.4$-$2.5\mK across the whole Q-band at this spectral resolution. 

Besides the Yebes 40m data, we have also analysed data obtained with the IRAM 30m radio telescope (Granada, Spain), as a result of projects 058-21 (PI: Rivilla) and  155-24 (PI: San Andr{\'e}s).  
The observations on project 058-21 spanned three night runs in September 2021 for a total time on source of $\sim$10.5\hrs, and covered the lower part of the 3\mm band ($\sim$72$-$103\GHz). The observations on project 155-24 were gathered during eight days between April and June 2025 achieving a total time on source of $\sim$15\hrs, and they covered the upper part of 3\mm band ($\sim$84$-$116.6\GHz). In both cases, we used the same On and Off positions as indicated above for the Yebes 40m observations, with the HPBW of the IRAM 30m telescope varying between $\sim$21$''$ to $\sim$34$''$ throughout the 3\mm window. The spectra were smoothed to $\sim$609\kHz ($\sim$1.57$-$2.54\kms in the range covered), and we achieved rms values from $\sim$0.6$-$4.6\mK along the $\sim$72$-$115\GHz range (rising up to $\sim$18\mK at the high frequency border at $>$115\GHz).

The line intensity of the spectra was directly measured in antenna temperature ($T_\text{A}^*$) units, since the molecular emission towards G+0.633, as in the case of G+0.693, is found to be extended over the primary beam of the telescope (San Andrés, in prep.). A more detailed description on all these observations and data reduction process will be provided in San Andrés et al., in prep.

\begin{figure*}
\centering
\includegraphics[width=\textwidth]{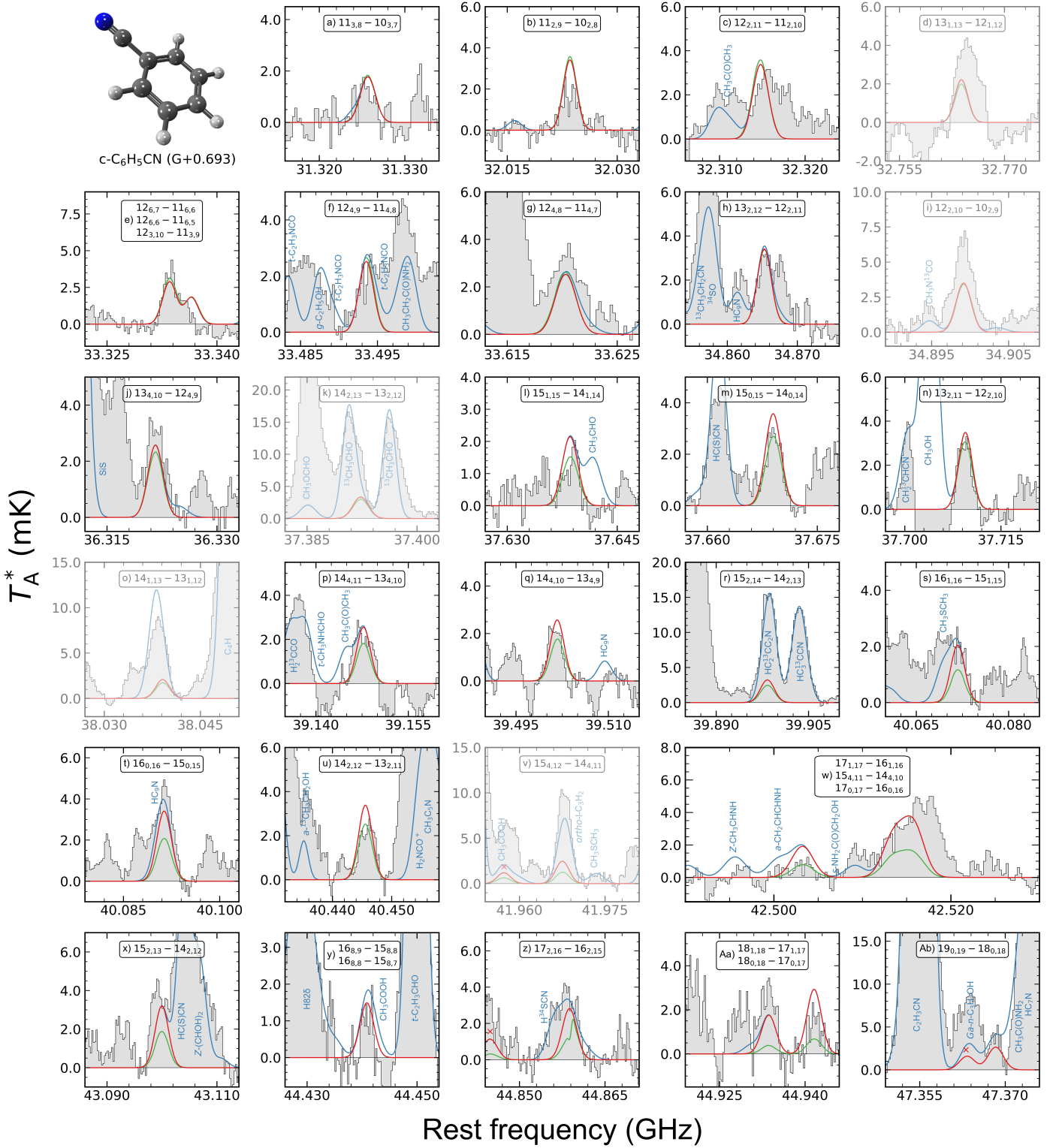}
\caption{Selected \benzo transitions identified towards the G+0.693 molecular cloud (spectroscopic details given in Table~\ref{tab:transitions}). The black histogram and grey-shaded areas indicate the observational spectra, while the red and blue solid lines represent the best LTE fit obtained for this molecule, and the aggregated emission profile considering all the molecules identified in this cloud (indicated by blue labels), respectively. The green line shows the best non-LTE prediction carried out for \benzo adopting: $n_{\ce{H2}} = 5\times10^{3}\, {\rm cm}^{-3}$, $T_\text{kin} = 100$\K, $N = (8.0 \pm 0.5)\times10^{12}\, \text{cm}^{-2}$ and $\text{FWHM} = 20$\kms, with the exception of transitions in panels y) and Ab) for which collisional data is missing. Panel labels indicate the \benzo rotational transitions being plotted using the common notation for asymmetric tops: $J_{K_\text{a}K_\text{c}}$. 
The transitions shown here are those used in the LTE analysis for \benzo in G+0.693, with the exception of: $i)$ those appearing in the faded panels, employed in the LTE fitting for the other cloud (G+0.633, see Fig.~\ref{fig:G+0.633_benzo_detection}), and $ii)$ those marked with a $\times$ symbol, excluded due to significant blending. 
The molecular structure of \benzo is depicted in the upper left side of the figure (C atoms in dark-grey, N atoms in blue and H atoms in white).} 
\label{fig:G+0.693_benzo_detection} 
\end{figure*}

\begin{figure*}[ht!]
    \centering
\includegraphics[width=\textwidth]{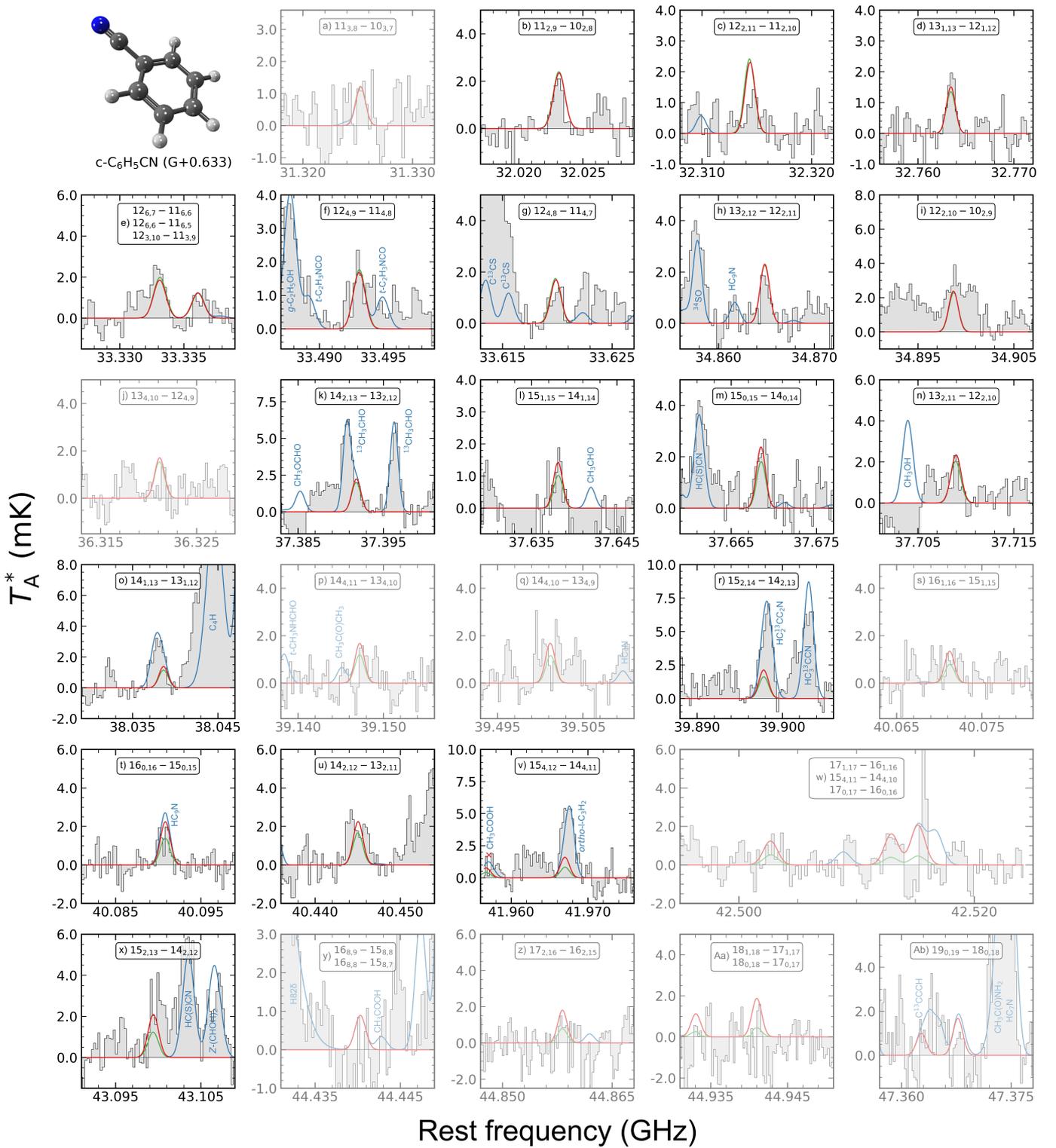}   
\caption{Same as Fig. \ref{fig:G+0.693_benzo_detection} but for G+0.633.
The transitions shown here are those that were used to perform the LTE analysis of \benzo in G+0.633, excluding those appearing in the faded panels due to the low SNR (which were used for the LTE fitting of \benzo in G+0.693 where they are clearly detected, see Fig.~\ref{fig:G+0.693_benzo_detection}), and the one marked in panel v) with a $\times$ symbol. The green lines depict the best non-LTE model achieved for G+0.633 (with the exception of panels y) and Ab) lacking collisional data), computed under $n_{\ce{H2}} = 5\times10^{3}\, {\rm cm}^{-3}$ and $T_\text{kin} = 60$\K and considering $N = (2.60 \pm 0.13)\times10^{12}\, \text{cm}^{-2}$, $\text{FWHM} = 10$\kms.} 
    \label{fig:G+0.633_benzo_detection} 
\end{figure*}

\begin{table*}
\centering
\tabcolsep 1.5pt
\caption{List of observed transitions of benzonitrile. We provide the transition frequencies (with their corresponding uncertainties associated to the last digits indicated in brackets), quantum numbers (plus an "o" or "p" label whether the transition is \textit{ortho} or \textit{para}, respectively), base 10 logarithm of the integrated intensity at 300 K (log $I$), Einstein coefficients ($A_{\rm ul}$), the values of the lower and upper level energy of each transition ($E_{\rm l}$ and $E_{\rm u}$), in cm$^{-1}$ and K, respectively. The last column gives the information about the species whose transitions are partially blended with benzonitrile lines.}
\begin{tabular}{ l c c  c c c c c }
\hline
Frequency & Transition & log $I$ & log $A_{\rm ul}$ & $E_{\rm l}$ & $E_{\rm u}$  & Panel  & Blending$^a$ (G+0.693 / G+0.633) \\
(GHz) &   & (nm$^2$ MHz) & (s$^{-1}$) & (cm$^{-1}$)  & (K) &  (Figs.~\ref{fig:G+0.693_benzo_detection}\&\ref{fig:G+0.633_benzo_detection}) & \\
\hline
31.3253330(9)$^{b}$ & $11_{3,8}-10_{3,7}$ p & -5.437 & -5.4908 & 6.4 & 10.7 &  a) & Unblended / Low SNR \\
32.023212(1)$^{b,c}$ & $11_{2,9}-10_{2,8}$ o & -5.175 & -5.4426 & 5.8 & 9.9 &  b) & Unblended / Unblended \\
32.314450(1)$^{b,c}$ & $12_{2,11}-11_{2,10}$ o & -5.131 & -5.4297 & 6.6 & 11.0 &  c) & Blended with U / Unblended \\
32.7634963(12)$^{b,c}$ & $13_{1,13}-12_{1,12}$ p & -5.297 & -5.4003 & 6.8 & 11.3 &  d) & Heavily blended with U / Slightly blended with U \\
33.333091(1)$^{b}$ & $12_{6,7}-11_{6,6}$ o & -5.224 & -5.4997 & 11.2 & 17.8 &  e) & Slightly blended with U / Slightly blended with U \\
33.333274(1)$^{b}$ & $12_{6,6}-11_{6,5}$ o & -5.224 & -5.4997 & 11.2 & 17.8 &  e) & Slightly blended with U / Slightly blended with U \\
33.336164(1)$^{b,c}$ & $12_{3,10}-11_{3,9}$ p & -5.342 & -5.4033 & 7.4 & 12.2 & e) & Slightly blended with U / Unblended \\
33.493181(1)$^{b}$ & $12_{4,9}-11_{4,8}$ o & -5.141 & -5.4197 & 8.4 & 13.7 & f) & Slightly blended with U / Slightly blended with U \\
33.619950(1)$^{b}$ & $12_{4,8}-11_{4,7}$ o & -5.137 & -5.4147 & 8.4 & 13.7 & g) & Slightly blended with U / Unblended \\
34.864904(1)$^{b,c}$ & $13_{2,12}-12_{2,11}$ o & -5.031 & -5.3278 & 7.7 & 12.7 & h) & Slightly blended with U / Unblended \\
34.8987676(11)$^{b,c}$ & $12_{2,10}-11_{2,9}$ o & -5.063 & -5.3271 & 6.9 & 11.6 & i) & Blended with U / Blended with U \\
36.321221(1)$^{b,c}$ & $13_{4,10}-12_{4,9}$ o & -5.030 & -5.3049 & 9.5& 15.4 & j) & Slightly blended with U / Low SNR \\
\multirow{2}{*}{37.3919131(11)$^{b,c}$} & \multirow{2}{*}{$14_{2,13}-13_{2,12}$ o} & \multirow{2}{*}{-4.940} & \multirow{2}{*}{-5.2343} & \multirow{2}{*}{8.8} & \multirow{2}{*}{14.5} & \multirow{2}{*}{k)} & Heavily blended with \ce{^{13}CH3CHO} / \\
&&&&&&& Blended with \ce{^{13}CH3CHO} \\
37.6381687(13)$^{b,c}$ & $15_{1,15}-14_{1,14}$ p & -5.119 & -5.2169 & 9.1 & 14.8 &  l) & Unblended / Unblended \\
37.6685672(13)$^{b,c}$ & $15_{0,15}-14_{0,14}$ o & -4.896 & -5.2158 & 9.1 & 14.8 &  m) & Unblended / Unblended \\
37.7090476(12)$^{b,c}$ & $13_{2,11}-12_{2,10}$ o & -4.962 & -5.2237 & 8.1 & 13.4 &  n) & Unblended / Unblended \\
38.0386987(12)$^{b,c}$ & $14_{1,13}-13_{1,12}$ p & -5.145 & -5.2100 & 8.7 & 14.4 &  o) &  Heavily blended with U / Blended with U \\ 
39.1473052(11)$^{b,c}$ & $14_{4,11}-13_{4,10}$ o & -4.929 & -5.1999 & 10.7 & 17.3 &  p) & Unblended / Low SNR \\
39.5012554(11)$^{b,c}$ & $14_{4,10}-13_{4,9}$ o & -4.921 & -5.1881 & 10.7 & 17.3 &  q) & Unblended / Low SNR \\
\multirow{2}{*}{39.8978884(12)$^{b,c}$} & \multirow{2}{*}{$15_{2,14}-14_{2,13}$ o}& \multirow{2}{*}{-4.855} & \multirow{2}{*}{-5.1479} & \multirow{2}{*}{10.1} & \multirow{2}{*}{16.4}  & \multirow{2}{*}{r)} & Heavily blended with \ce{HC2^{13}CC2N} / \\
&&&&&&& Heavily blended with \ce{HC2^{13}CC2N} \\
40.0712945(14)$^{b,c}$ & $16_{1,16}-15_{1,15}$ p & -5.038 & -5.1341 & 10.3 & 16.8 &  s) & Blended with U / Low SNR \\
40.0909107(14)$^{b,c}$ & $16_{0,16}-15_{0,15}$ o & -4.816 & -5.1335 & 10.3 & 16.8 &  t) & Slightly blended with \ce{HC9N} and U / Unblended \\
40.4451233(13)$^{b,c}$ & $14_{2,12}-13_{2,11}$ o & -4.871 & -5.1308 & 9.3 & 15.3 &  u) & Unblended / Slightly blended with U \\
\multirow{2}{*}{41.9670785(12)$^{b,c}$} & \multirow{2}{*}{$15_{4,12}-14_{4,11}$ o} & \multirow{2}{*}{-4.837} & \multirow{2}{*}{-5.1034} & \multirow{2}{*}{12.0} & \multirow{2}{*}{19.3} & \multirow{2}{*}{v)} & Heavily blended with l-\ce{C3H2} and U / \\
&&&&&&& Heavily blended with l-\ce{C3H2} \\
42.5027671(15)$^{b,c}$ & $17_{1,17}-16_{1,16}$ p & -4.964 & -5.0564 & 11.6 & 18.8 &   w) & Unblended / Low SNR \\
42.5130495(12)$^{b,c}$ & $15_{4,11}-14_{4,10}$ o & -4.825 & -5.0864 & 12.1 & 19.4 &  w) & Blended with U / Low SNR \\
42.5153031(15)$^{b,c}$ & $17_{0,17}-16_{0,16}$ o & -4.742 & -5.0560 & 11.6 & 18.8 &  w) & Blended with U / Low SNR \\
43.0995165(14)$^{b,c}$ & $15_{2,13}-14_{2,12}$ o & -4.789 & -5.0468 & 10.7 & 17.4 &  x) & Slightly blended with \ce{HC(S)CN} / Blended with U \\
44.4403894(14)$^{b}$ & $16_{8,9}-15_{8,8}$ o & -4.869 & -5.1205 & 20.2 & 31.2  & y) & Slightly blended with \ce{CH3COOH} / Low SNR \\
44.4403936(14)$^{b}$ & $16_{8,8}-15_{8,7}$ o & -4.869 & -5.1205 & 20.2 & 31.2 &  y) & Slightly blended with \ce{CH3COOH} / Low SNR \\
44.8582981(13)$^{b}$ & $17_{2,16}-16_{2,15}$ o & -4.704 & -4.9922 & 12.8 & 20.6 &  z) & Blended with \ce{H^{34}SCN} / Low SNR \\
44.9331451(16)$^{b}$ & $18_{1,18}-17_{1,17}$ p & -4.893 & -4.9831 & 13.1 & 21.0 &  Aa) & Blended with U / Low SNR \\
44.9410905(16)$^{b,c}$ & $18_{0,18}-17_{0,17}$ o & -4.671 & -4.9828 & 13.1 & 21.0 &  Aa) & Unblended / Low SNR \\
47.3678123(17)$^{b,c}$ & $19_{0,19}-18_{0,18}$ o & -4.605 & -4.9136 & 14.6 & 23.2 &  Ab) & Slightly blended with \ce{CH3CONH2} / Low SNR \\
\hline 
\end{tabular}
\label{tab:transitions}
{\\ (a) ``U'' refers to blendings with emission from an unknown (not identified) species. On the other hand, the tag ``Low SNR'' corresponds to lines traced below a 3$\sigma$ detection threshold (establishing a \scalebox{0.85}{\usym{2613}} in the source column accordingly), with $\sigma$ delineating the observational rms at the frequency of the line.}
{\\ (b) \benzo Transition identified by \citet{cernicharo2021_ethynylcyclopentadiene} towards TMC-1 (see their appendix C).}
{\\ (c) \benzo Transition identified by \citet{cernicharo2023_maps} towards TMC-1.}

\end{table*}

\section{Analysis and Results}\label{sec:analysis and results}

\subsection{Identification and LTE analysis}

We have performed the identification and derivation of the molecular abundances of \benzo towards both G+0.693 and G+0.633 through a Local Thermodynamic Equilibrium (LTE) analysis, which assumes that the population of the different energy levels of the molecules can be described by a unique excitation temperature ($T_{\rm ex}$). The densities of the two targeted molecular clouds are intermediate, $\sim$10$^{4}-10^{5}$ cm$^{-3}$ (\citealt{zeng2020,Colzi2021,Colzi2024}, and San Andrés et al., in prep.), producing subthermal excitation of the molecules. This translates into excitation temperatures in a typical range of 3$-$20 K (e.g. \citealt{Zeng2018,Jimenez-Serra2020,Rivilla2022_nitriles,SanAndres2023}) that are significantly lower than the kinetic temperatures of the sources ($T_{\rm kin}$=60$-$100 K). However, the population of the molecules can still be well reproduced by an LTE model with a single $T_{\rm ex}$ because these reach a situation of quasi thermalization. This behaviour is explained in detail in \citet{Goldsmith1999}. 
Moreover, the LTE analysis also provides a good estimate of the molecular column density, as demonstrated previously in other works targetting G+0.693 in which LTE and non-LTE analyses have been compared (e.g., \citealt{Colzi2021,Colzi2024, Massalkhi2023}, and Sanz-Novo et al., in prep).  

We have conducted the analysis of \benzo towards both clouds by using the version from 2024 December 11 of the Spectral Line Identification and Modelling (SLIM) tool within the \textsc{MADCUBA}\footnote{Madrid Data Cube Analysis on ImageJ is a software developed at the Centro de Astrobiología (CAB) in Madrid: \url{https://cab.inta-csic.es/madcuba/}} package \citep{Martin2019}. To do so, we uploaded into it the entry 103501 (dated November 2009) included within the Cologne Database for Molecular Spectroscopy (CDMS, \citealt{endres2016}), which contains \benzo transition frequencies measured by different laboratory works (\citealt{casado1971,fliege1981,vormann1988,wlodarczak1989,wohlfart2008}). The dipole moment of \benzo is relatively large, 4.51 D, and was determined by \citet{wohlfart2008}. Within SLIM, we then generated a synthetic spectrum under the assumption of LTE conditions, which we fitted to the observational ones by simultaneously considering the contribution from the $>$145 molecules so far identified towards G+0.693 (e.g., \citealt{Sanz-Novo2025}) and the $>$100 species in G+0.633 (San Andr{\'e}s et al., in prep.). The physical parameters describing the molecular emission are the total column density of the molecule ($N$), the excitation temperature ($T_\text{ex}$), the local standard of rest velocity ($v_\text{LSR}$), and the full width at half maximum ($\text{FWHM}$).

Figs.~\ref{fig:G+0.693_benzo_detection} and \ref{fig:G+0.633_benzo_detection} show all \benzo transitions identified towards G+0.693 and G+0.633, respectively, that are either unblended or present certain blending with other molecular species, but whose combined contributions closely reproduce the observed spectrum. We note that some transitions appear unblended in G+0.633, which exhibits narrower linewidths (see below), but have a higher level of contamination in G+0.693. Moreover, some of the transitions clearly detected in G+0.693 (which have brighter line intensities) are close to the current noise limits in G+0.633. 
All these contaminated and low signal-to-noise transitions are shown with transparency in Figs.~\ref{fig:G+0.693_benzo_detection} and \ref{fig:G+0.633_benzo_detection} for consistency, but were excluded in the LTE fitting. 
We also note that all of the transitions shown in both figures were already identified as \benzo spectral lines by \citet{cernicharo2021_ethynylcyclopentadiene,cernicharo2023_maps}. Their spectroscopical details are given in Table~\ref{tab:transitions}.

To obtain the best LTE model for \benzo in both clouds, we run the tool \textsc{SLIM-AUTOFIT} to fit all the transitions depicted in Figs.~\ref{fig:G+0.693_benzo_detection} and \ref{fig:G+0.633_benzo_detection}, after excluding those depicted with transparency, as mentioned before. \textsc{SLIM-AUTOFIT}  provides the best non-linear least-squares LTE fit to the data using the Levenberg-Marquardt algorithm (see \citealt{Martin2019} for a detailed description of the formalism). 
In the case of G+0.633, we had to fix both the $T_\text{ex}$ and $\text{FWHM}$ to allow for convergence, which were assumed to be 10\K and 10\kms, respectively, in accordance to those derived for many of the nitriles identified so far (San Andr{\'e}s et al., in prep.). In the same way, we fixed the $\text{FWHM}$ when performing the fitting for G+0.693, adopting a value of 20\kms as found for other \ce{N}-bearing species detected in this cloud (e.g., \citealt{Rivilla2022_nitriles}; \citealt{SanAndres2024}). Keeping these parameters fixed, we derived: $N = (8.0 \pm 0.5)\times10^{12}\, \text{cm}^{-2}$, $T_\text{ex} = 10.7 \pm 1.8$\K and $v_\text{LSR} = 65.7 \pm 0.7$\kms for G+0.693, and $N = (2.60 \pm 0.13)\times10^{12}\, \text{cm}^{-2}$ and $v_\text{LSR} = 50.0 \pm 0.3$\kms for G+0.633.

To derive the molecular abundance of \benzo relative to \ce{H2} towards G+0.693, we employed $N$(H$_2$) = 1.35$\times$10$^{23}$ cm$^{-2}$, as derived from \ce{C^{18}O} by \citet{Martin2008}. In the case of G+0.633, we 
used $N({\rm H}_2) = (6.0 \pm 1.2) \times 10^{22}$ cm$^{-2}$, based on the analysis of \ce{C^{18}O}, \ce{C^{17}O}, \ce{^{13}C^{18}O} and \ce{^{13}C^{17}O} isotopologues detected in this cloud (San Andrés et al., in prep.).
Using these values, we obtained that the molecular abundances of \benzo towards G+0.693 and G+0.633 are quite similar: (6$\pm$1)$\times$10$^{-11}$ and (4.3$\pm$0.9)$\times$10$^{-11}$, respectively.

\begin{figure}
    \centering
    \includegraphics[width=\linewidth]{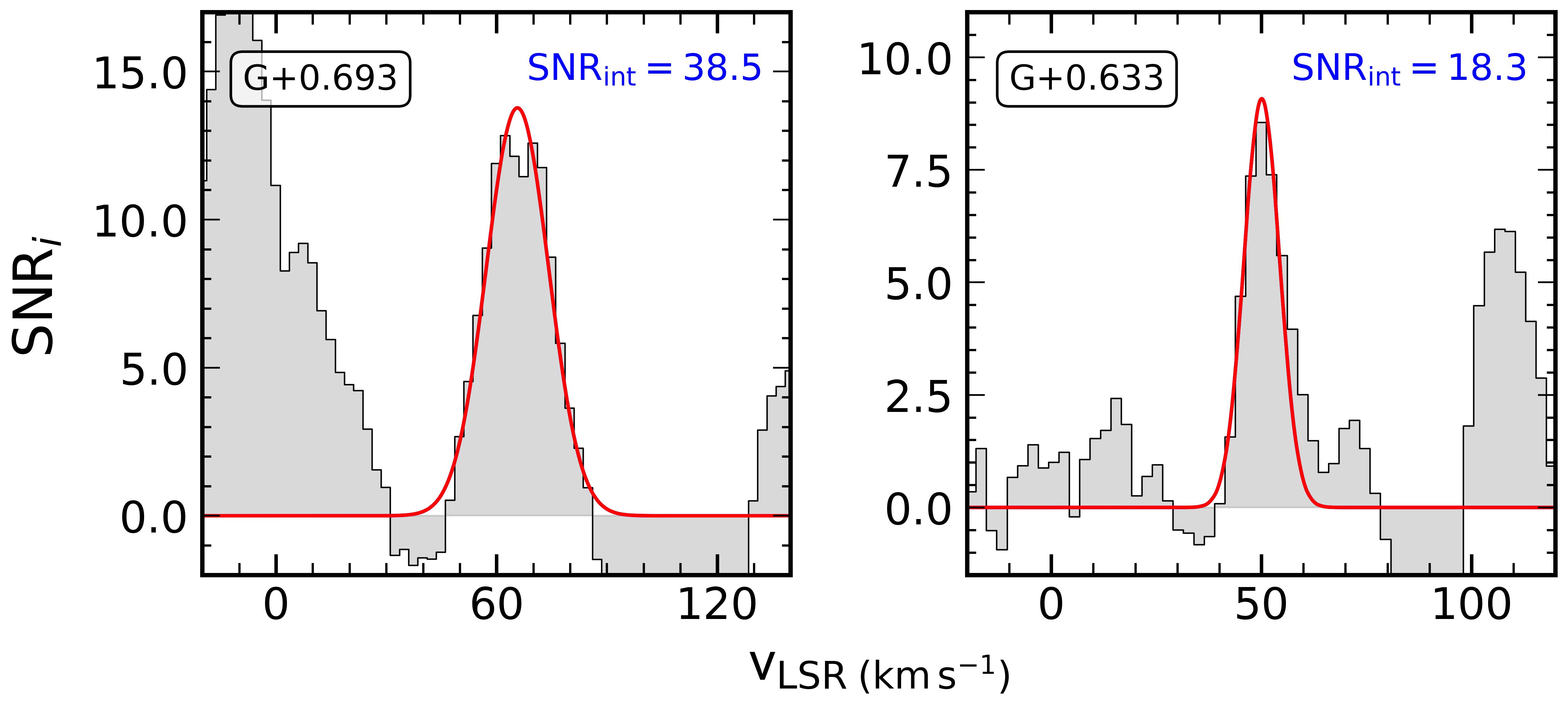}
    \caption{Stacking in G+0.693 (left) and G+0.633 (right), centered at the velocity of each source). To compute it, we used the 11 most unblended transitions shown in Fig.~\ref{fig:G+0.693_benzo_detection} for G+0.693 (depicted in panels a, b, e, j, l-n, p, q, u and y); and the 9 most unblended transitions appearing in Fig.~\ref{fig:G+0.633_benzo_detection} for G+0.633 (panels b, d, e, g, h, l-n and t). The rms of the stacked spectrum is 0.18\mK in the case of G+0.693 and 0.2\mK for G+0.633. Both plots are scaled in terms of the SNR for individual channels (SNR$_i$), while we show in blue at the top right corner the SNR in integrated intensity of c-\ce{C6H5CN} stacked profile. The signals at velocities different to those of the sources are due to transitions of other molecular species that are present in the individual spectrum used to perform the stacking.}
    \label{fig:stackings} 
\end{figure}

Although the line-by-line identification of \benzo towards both clouds is robust thanks to the detection of multiple transitions, to quantify the detection significance we also performed a stacking analysis
(see \citealt{Loomis2021} for a detailed description of this kind of technique).
We used the stacking procedure implemented in MADCUBA-SLIM, as done in \citet{Sanz-novo2025_mf}.
We selected the most unblended transitions towards each cloud, which are 11 and 9 for G+0.693 and G+0.633, respectively. These transitions are indicated in panels a), b), e), j), l-n), p), q), u) and y) in Fig.~\ref{fig:G+0.693_benzo_detection} for G+0.693, and panels b), d), e), g), h), l-n) and t) in Fig.~\ref{fig:G+0.633_benzo_detection} for G+0.633. 
After subtraction of the modeled contribution of other species from the observed spectrum, we stacked the remaining spectra in velocity scale. The individual spectra around the different transitions are weighted according to the intensity modeled by the best LTE fit of \benzo. We show the resulting stacked spectra in Fig. \ref{fig:stackings}, in which the y-axis indicates the signal-to-noise ratio per channel (SNR$_i$). The values at the line peaks are $\sim$12 and $\sim$8 for G+0.693 and G+0.633, respectively. We note that these values are indeed lower limits of the detection level, since the emission lines have a Gaussian shape. To account for this we also calculated the integrated signal-to-noise (SNR$_{\rm int}$) of the stacked spectra integrated with SNR$_{\rm int}$ = $\int$ $T$$\mathrm{_A^*}$d$v$ / ( rms $\times$ $\sqrt{\updelta v \times \mathrm{FWHM}}$), where rms is the measured root mean square noise of the stacked spectra, $\updelta$$v$ is the velocity resolution of the spectra, and FWHM indicates the full width at half maximum of the Gaussians. The derived values of SNR$_{\rm int}$ are 38.5 and 18.3 for G+0.693 and G+0.633, respectively. 

\subsection{Non-LTE analysis}

We have also performed a non-LTE analysis of \benzo to explore the effect of the gas $T_{\rm kin}$ and volume density ($n_{\ce{H2}}$) on the expected line intensities, using the collisional coefficients of \benzo with He atoms recently computed by \citet{khalifa2024}. In this work the potential energy surface (PES) corresponding to the interaction of \benzo with helium was computed along with the collisional rates for all rotational levels up to $J=9$ and $T_{\rm kin}$ = 40\K. Since higher rotational transitions have been detected towards G+0.693 and G+0.633 (up to $J$=19, see Table \ref{tab:transitions}), and these two sources have higher $T_\text{kin}$, we have extended the \citet{khalifa2024} analysis to 100\K and $J=19$\footnote{We note that two of the detected transitions, shown in panels y) and Ab) of Figs. \ref{fig:G+0.693_benzo_detection} and \ref{fig:G+0.633_benzo_detection}, were not included in the calculations, because the energy of the upper levels  energies exceed the threshold considered to allow convergence in the scattering calculations. For this reason, the non-LTE predictions for these transitions (green curves) are not shown in Figs. \ref{fig:G+0.693_benzo_detection} and \ref{fig:G+0.633_benzo_detection}.}. Additional details are given in Appendix \ref{sec:appendix-rates}.
We note that the rotational levels of \benzo are divided into {\it ortho} and {\it para} due to the permutation symmetry of the hydrogen atoms and their nuclear spin $I$ = 1/2. The nuclear spin symmetry prevents the interconversion of the {\it ortho} and {\it para} levels through radiative or inelastic collisional processes. Consequently, the scattering calculations are performed separately for each nuclear spin species. In addition, the non-LTE predictions were scaled by a factor of 1.4 to account for the reduced mass difference between \ce{He} and \ce{H2} (the major collider in the ISM).

Using the newly calculated rates, we run the statistical equilibrium radiative transfer code 
\textsc{Radex}\footnote{https://home.strw.leidenuniv.nl/~moldata/radex.html} (\citealt{vandertak2007}). We performed a grid of non-LTE models by varying the values of $n_{\ce{H2}}$ and $T_\text{kin}$ within the ranges derived in these clouds (\citealt{Zeng2018,zeng2020,Colzi2022}; and San Andrés et al. in prep). In particular, we run models for $n_{\ce{H2}}$ of 5$\times10^{3}$, $10^{4}$ and 2.5$\times10^{4}$ cm$^{-3}$; and for $T_\text{kin}$ of 60, 100 and 140\K \footnote{The calculation of the collisional rates was done up to 100 K, therefore being those at 140 K (which was the highest $T_\text{kin}$ derived by \citet{Zeng2018} towards G+0.693) extrapolated. This was done to test if significant differences with respect to 100 K are seen.}. For all non-LTE models we adopted as input the column density derived from the LTE analysis, namely $N = (8.0 \pm 0.5)\times10^{12}\, \text{cm}^{-2}$ for G+0.693, and $N = (2.60 \pm 0.13)\times10^{12}\, \text{cm}^{-2}$ for G+0.633. 

To enable a direct comparison between the non-LTE results, where the \textit{para} and \textit{ortho} levels of \benzo are treated separately, and the LTE results based on the CDMS entry, which incorporates nuclear spin statistics in the intensity calculations (i.e. assuming an \textit{ortho}:\textit{para} ratio of 5:3), we adopted the same 5:3 ratio in the column density used as input for the \textsc{Radex} computations, i.e., $N$(total) = 0.625 $\times$ $N$({\it ortho}) + 0.375 $\times$ $N$({\it para}). We note that \textsc{Radex} recomputes the rotational partition function for the molecule using the rotational energy levels included in the input (.dat) file.
We found that the non-LTE prediction for the line profiles of all \benzo transitions analysed in this work are not particularly sensitive to $T_\text{kin}$ (because the rate coefficients at these high temperatures have only a weak temperature-dependence), but to $n_{\ce{H2}}$. In this regard, we found that the lowest $n_{\ce{H2}}$ values explored, 5$\times 10^{3} \; {\rm cm}^{-3}$, overall explained better the observed \benzo spectrum for both clouds, although the higher $E_{\rm up}$ transitions are slightly better reproduced with the intermediate $\ce{H2}$ densities of 10$^{4} \; {\rm cm}^{-3}$. This may suggest a possible density structure linked to the \benzo emitting gas. The predictions for \benzo lines profiles from the best non-LTE models found are displayed in green in Figs.~\ref{fig:G+0.693_benzo_detection} and ~\ref{fig:G+0.633_benzo_detection}, corresponding to: 
$n_{\ce{H2}} = 5\times10^{3}\, {\rm cm}^{-3}$ and $T_\text{kin} = 100$\K for G+0.693; whereas 
$n_{\ce{H2}} = 5\times10^{3}\, {\rm cm}^{-3}$ and $T_\text{kin} = 60$\K for G+0.633. 

\begin{table}
\tabcolsep 1.95pt
\centering
\caption{Derived physical parameters of benzonitrile detected towards G+0.693-0.027 and G+0.633-0.0604. The parameters without uncertainties were fixed for the fit.}
\begin{tabular}{ c c c c c c   }
\hline
Cloud  & $N$ &  $T_{\rm ex}$ & $v$$_{\rm LSR}$ & FWHM  & $\chi$    \\
  &($\times$10$^{12}$ cm$^{-2}$) & (K) & (km s$^{-1}$) & (km s$^{-1}$) & ($\times$10$^{-11}$)      \\
\hline
G+0.693 & 8.0$\pm$0.5 & 10.7$\pm$1.8 & 65.7$\pm$0.7 & 20.0 &   6$\pm$1 $^a$ \\
G+0.633 & 2.60$\pm$0.13 & 10.0 & 50.0$\pm$0.3 & 10.0 & 4.3$\pm$0.9 $^b$ \\
\hline
\\[-7pt]
\multicolumn{6}{p{.45\textwidth}}{$^a$ Calculated by adopting $N_{\rm H_2}$ = 1.35$\times$10$^{23}$ cm$^{-2}$, from \citet{Martin2008}, assuming an uncertainty of 15\% of its value; $^b$ Computed by adopting $N_{\rm H_2}$ = (0.60$\pm$0.12)$\times$10$^{23}$ cm$^{-2}$ from San Andrés et al., in prep.}
\end{tabular}
\label{tab:comparison}
\end{table}

\section{Discussion}\label{sec:discussion}

The detections of \benzo towards two CMZ molecular clouds have relevant implications about the distribution of aromatics throughout the Galaxy, and also about its formation and survival in the ISM.
First, this work confirms that \benzo is widespread in the Galaxy, extending its presence from the vicinity of our Solar System ($d<$ 500 pc) where it had been previously detected, to the center of our Galaxy located at 8.2 kpc (\citealt{gravity2019}) from Earth. 

Secondly, it confirms that \benzo can be present in significantly harsher environments than that of cold clouds.
Unlike the low temperatures in Galactic disk molecular clouds ($\sim$10 K), G+0.693 and G+0.633 exhibit much higher kinetic temperatures of 70$-$140 K (\citealt{Zeng2018}, and San Andr\'es, in prep.), while having low dust temperatures of $\sim$20 K (\citealt{battersby2025}).
Moreover, similarly to other molecular clouds in the CMZ, their chemistry is also affected by shocks, as demonstrated by bright emission of typical shock tracers such as HNCO and SiO (\citealt{zeng2020}, and San Andr\'es, in prep.), and levels of cosmic-ray ionisation rates several orders of magnitude higher than those typically found in the Galactic disk (\citealt{Rivilla2022_PO+,Sanz-Novo2024a}), which also produce an intense radiation field of secondary UV photons.  
These different environmental conditions might influence the aromatic chemistry, making the CMZ a good chemical laboratory. 
In the following sections we will discuss the implications of the detections presented in this work on the presence, formation and relevance of \benzo and of larger aromatics in the ISM.

\subsection{Benzonitrile across different interstellar environments and comparison with \ce{HC7N}}
\label{sec:comparison-hc7n}

\begin{figure}
    \centering
    \includegraphics[width=\linewidth]{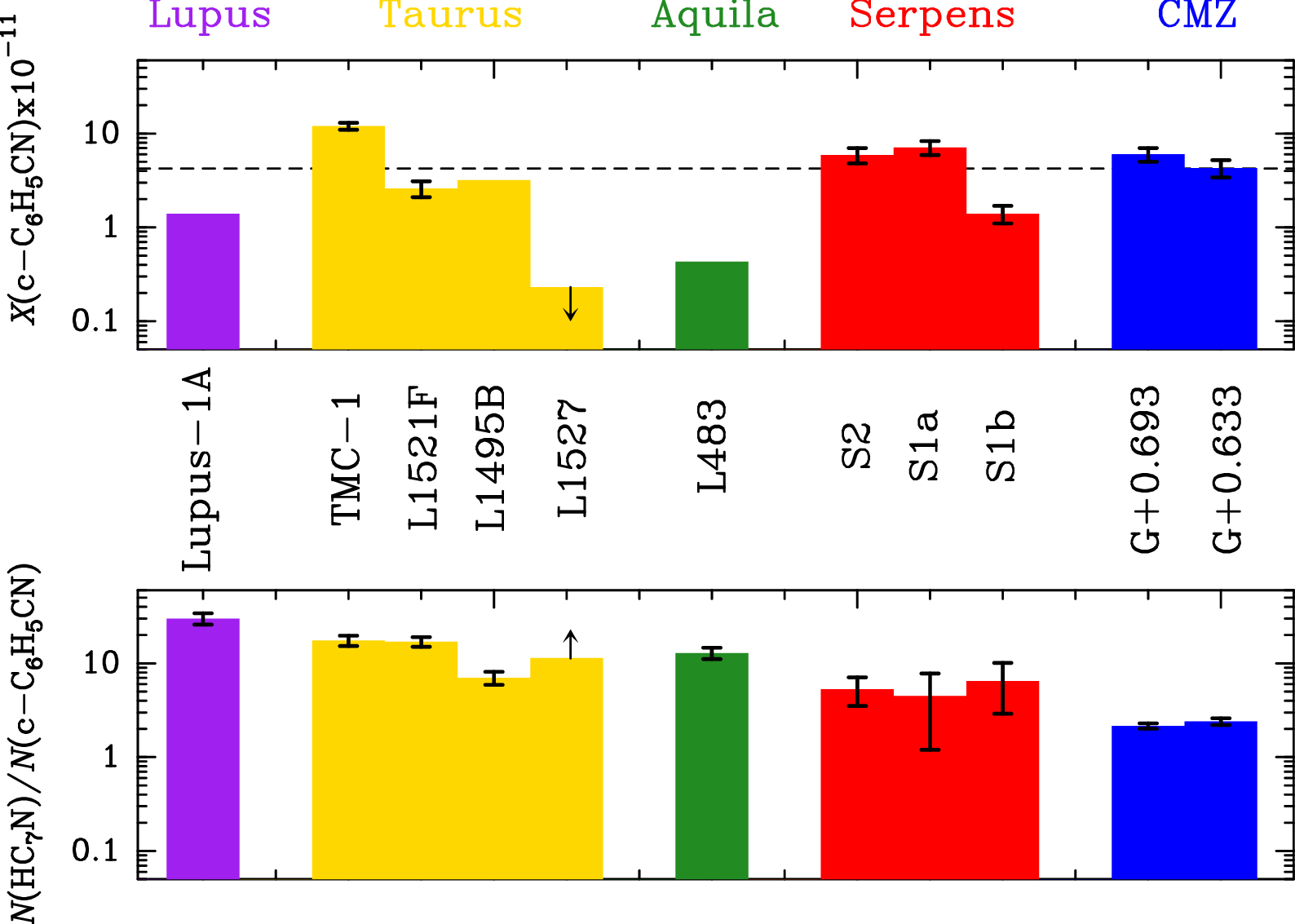}
    \vskip 3mm
    \hspace{-1mm}
    \includegraphics[width=1\linewidth]{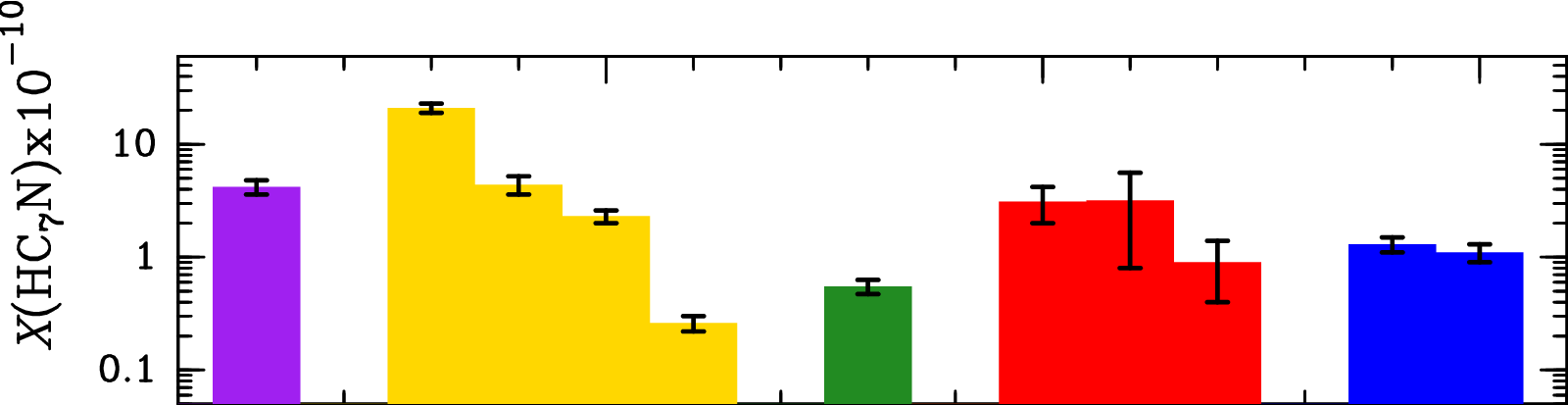}
    \caption{{\it Upper panel}: molecular abundance of \benzo towards galactic cold clouds and CMZ molecular clouds. The clouds from the same galactic complex have been grouped and denoted with the same color. The arrow indicates an upper limit. The dashed horizontal line indicate the average value of cold clouds, without considering the upper limit. {\it Middle pannel:} Same as upper panel, but for the molecular ratio \ce{HC7N}/\benzo. The arrow indicates a lower limit. {\it Lower panel:} Same as in previous panels, but for the \ce{HC7N} abundances.}
    \label{fig:histo} 
\end{figure}

We show in the upper panel of Fig. \ref{fig:histo} the  molecular abundances of \benzo derived towards the CMZ clouds compared with those previously found towards cold molecular clouds located in several galactic complexes that include Lupus, Taurus, Aquila, and Serpens, from \citet{burkhardt2021,cernicharo2021_ethynylcyclopentadiene,agundez2023a} (see also Table \ref{tab:comparison}). The abundances in the CMZ, (4.3$-$6.0)$\times$10$^{-11}$, are very similar (within a factor of 2) to those reported in L1495B (Taurus) and S2 and S1a (Serpens), and strikingly coincident with the average value of the abundance found in cold clouds, which is $\sim$4.8$\times$10$^{-11}$ (indicated with a black horizontal dashed line in Fig. \ref{fig:histo}). This latter average does not include L1527 (in Taurus), where \benzo was not detected and only an upper limit was reported. Interestingly, this source contains a protostar within the cloud (\citealt{Sakai2008}), which is also the case of the other source with lowest abundance of \benzo, L483 (in Aquila; \citealt{Agundez2019}), as pointed out by \citet{agundez2023a}. This, and the lack of detection of \benzo so far in hot cores/corinos, might indicate that protostellar sources are environments prone to the destruction of \benzo.

We have also compared the column densities of \benzo found in all sources with that of \ce{HC7N}. For the CMZ clouds, we include the analysis in Appendix \ref{sec:appendix-hc7n}; while the values in the Galactic clouds are obtained from \citet{agundez2023a} (see references therein). We chose this species because it shares the same number of heavy atoms with \benzo, and it is usually used as a proxy of rich carbon-chain chemistry. Moreover, it has been proposed that carbon-chain 
molecules such as HC$_7$N might be precursors of PAHs (e.g. \citealt{guzman-ramirez2011}). 
\citet{agundez2023a} showed that for cold clouds the column densities of \benzo and \ce{HC7N} show a overall reasonable trend (their Fig. 3), which was interpreted as a possible chemical relation among aromatic cycles and carbon chains.   

In the middle panel of Fig. \ref{fig:histo} we show the \ce{HC7N}/\benzo ratio (see values in Table \ref{tab:comparison}) in the whole sample of sources with \benzo is detected, adding the two CMZ clouds. 
The \ce{HC7N}/\benzo ratio is far from being constant, presenting variations above one order of magnitude among sources. 
The value found in CMZ clouds is nearly the same within uncertainties (2.15-2.4) is lower than those derived in Galactic clouds, which spans from 4.5 to 30 (middle panel in Fig. \ref{fig:histo}). 
This high dispersion of the ratio seems to rule out a direct chemical link among these carbon-rich species.

Interestingly, despite this overall dispersion, the ratio seems to be relatively constant for the sources located in the same complex.
On average, the \ce{HC7N}/\benzo ratio decreases following this sequence: Lupus $>$ Taurus $>$ Aquila $>$ Serpens $>$ CMZ.
This suggests that local effects in the parental Galactic complexes might have a prevalence for triggering aromatic vs. carbon-chain chemistry, with the CMZ being the place of the Galaxy where aromatic chemistry is more dominant (lower \ce{HC7N}/\benzo ratio). 

Additionally, for G+0.693 we compared the abundance of \benzo derived in this work with that of the sulfur-bearing cycle 2,5-cyclohexadien-1-thione (2,5-CT; c-C$_6$H$_6$S), a structural isomer of thiophenol that has been recently detected towards this cloud \citep{Araki2026}. The two six-membered cyclic species exhibit comparable abundances, yielding a \benzo/2,5-CT abundance ratio of $1.2 \pm 0.1$. This similarity may suggest that pure aromatic hydrocarbons and also sulfur-containing cyclic species may share common or at least similar formation pathways and/or survivability in the extreme physical conditions of the CMZ.

\begin{table}
\tabcolsep 1.95pt
\centering
\caption{Molecular abundances of \benzo, and molecular ratios with respect to \ce{HC7N}, derived towards the CMZ molecular clouds G+0.693-0.027 and G+0.633-0.0604, and towards several Galactic cold clouds. Uncertainties are provided when available in the literature.}
\begin{tabular}{ c c c  c}
\hline
 Cloud  &  HC$_7$N / c-C$_6$H$_5$CN & $\chi$(c-C$_6$H$_5$CN)$^a$ & Region \\
    &   & ($\times$10$^{-11}$) &  \\
 \hline
 \multicolumn{4}{c}{CMZ molecular clouds} \\
 \hline
 G+0.693  & 2.15$\pm$0.14 & 6$\pm$1  & CMZ   \\
 G+0.633 & 2.4$\pm$0.2$^b$ & 4.3$\pm$0.9  & CMZ \\ 
\hline
 \multicolumn{4}{c}{Galactic cold clouds} \\
 \hline
  TMC-1$^c$  & 17.5$\pm$2.2 &   12$\pm$1 & Taurus \\ 
  MC27/L1521F$^d$  & 17$\pm$2   &   2.6$\pm$0.5 & Taurus \\
  L1495B$^e$   & 7.1$\pm$1.1   & 3.2   & Taurus \\ 
  L1527$^e$     &  $>$  11.4  & $<$ 0.23   & Taurus \\ 
  S2$^d$   & 5.3$\pm$1.8   &   5.9$\pm$1.1 & Serpens \\ 
  S1a$^d$     & 4.5$\pm$3.3  &   7.1$\pm$1.2  & Serpens\\ 
  S1b$^d$  & 6.5$\pm$3.6   &   1.39$\pm$0.25 & Serpens \\
  Lupus-1A$^e$    & 30$\pm$4   &  1.4  & Lupus \\ 
  L483$^e$    &  12.9$\pm$1.8  &  0.43  &  Aquila\\ 
\hline
\end{tabular}
\label{tab:comparison}
\label{tab:mol-ratios}
\begin{list}{}
\item $^{\mathrm{a}}$ The values for $N$(H$_2$) are taken from \citet{Martin2008} (G+0.693); San Andrés et al., in prep. (G+0.633); \citet{cernicharo2018} (TMC-1); \citet{burkhardt2021} (MC27/L1521F, S2, S1a, S1b); \citet{agundez2015} (L1495B, Lupus-1A); \citet{jorgensen2002} (L1527); and \citet{Agundez2019} (L483). For the uncertainties of the Galactic cold clouds we have considered those of $N$(\benzo), if available, and assumed a 15$\%$ for $N$(H$_{\rm 2}$).
\item$^{\mathrm{b}}$ Using the $N$(\ce{HC7N}) of the main velocity component (namely C1, see Appendix \ref{sec:appendix-hc7n}), which dominates the molecular emission of this species, and also corresponds to the c-\ce{C6H5CN} emission in this cloud. 
\item$^{\mathrm{c}}$ The \benzo and \ce{HC7N} column densities were taken from \citet{cernicharo2021_ethynylcyclopentadiene} and \citet{Cabezas2022}, respectively.
\item$^{\mathrm{d}}$ \benzo column densities taken from \citet{burkhardt2021}. The \ce{HC7N} column densities for S2, S1a and S1b were retrieved from \citet{Friesen2013}. 
The \benzo/\ce{HC7N} ratio is directly taken from \citet{burkhardt2021}.
\item$^{\mathrm{e}}$ \benzo and \ce{HC7N} column densities from \citet{agundez2023a}.
\end{list}
\end{table}

\subsection{Formation of \benzo (and larger aromatics) in the ISM}
\label{sec:discussion-chemistry}

It has been proposed that \benzo can be formed through the exothermic and barrierless gas-phase neutral-neutral reaction between the CN radical and benzene (\citealt{balucani2000,woon2006,trevitt2010}): CN + c-\ce{C6H6} $\rightarrow$ c-\ce{C6H5CN} + H. 
Laboratory measurements by \citet{cooke2020} found that the reaction rate does not show significant dependence with the temperature in the range between 10$-$100 K, indicating that it can occur in both temperature regimes. This agrees with the observational detection of \benzo both in low and high kinetic temperature molecular clouds. 

Similarly, the addition of the CN radical to larger cyclic rings has been proposed as the formation mechanism of the CN derivatives of PAHs detected towards TMC-1 (\citealt{Wenzel2024b}). In the case of G+0.693, analogous reactions involving also the CN radical interacting with other C-chain species have been also proposed as viable formation routes. For instance, the observed isomeric ratios for the \ce{C2H2N2} and \ce{C4H3N} isomeric families are well reproduced by the gas-phase reactions between the CN radical and \ce{CH2NH} (\citealt{SanAndres2024}), and CN with the unsaturated hydrocarbons of the \ce{C3H4} family (\citealt{Rivilla2022_nitriles}), respectively. 

Alternatively, \citet{cernicharo2021_ethynylcyclopentadiene} also considered the gas-phase reaction between CCH and 1-cyano-1,3-butadiene (\ce{CH2CHCHCHCN} or \ch{CH2(CH)3CN}).
Since rotational spectroscopy is available for the $Z$- and $E$-isomers of the latter species (CDMS entries 079502 and 079503; from \citealt{McCarthy2020}), we also searched for them towards G+0.693 and G+0.633 to test if they can be indeed viable gas-phase precursors of \benzo. None of them were detected, and we computed upper limits for their column densities. The abundances of the $Z$- and $E$-isomers are $\geq$ 4 and 9 times lower than \benzo in G+0.693, and $\geq$2.5 and 4.5 times lower  towards G+0.633, respectively, therefore ruling out their role as plausible precursors.

Overall, these results indicate that the most likely pathway to \benzo is directly from benzene. Thus, a key question naturally follows: how is c-\ce{C6H6} formed in the ISM in the first place? The formation of this first aromatic ring is a long-standing bottleneck in interstellar chemistry \citep{Jones2010,Kocheril2025,Loison2025}, with its formation pathways diverging between bottom-up and top-down mechanisms. The proposed bottom-up routes include:
(i) the formation of benzene in low-temperature acetylene (\ce{C2H2}) ices irradiated by energetic electrons, as shown in laboratory experiments by \citet{zhou2010};
(ii) the dissociative recombination of the c-\ce{C6H7+} ion (\citealt{mcewan1999,Woods2011,Agundez2021});
(iii) reaction networks involving 1,3-butadiene and radicals such as CCH or CN, leading to cyclopentadiene and subsequently benzene (e.g. \citealt{He2020,Caster2021}); and
(iv) the recombination of two propargyl radicals (\ce{C3H3}).

However, none of these mechanisms appears fully satisfactory. If benzene were efficiently produced by cosmic-ray processing of acetylene ices, its abundance would be expected to be higher in the CMZ—where cosmic-ray ionisation rates are elevated—than in cold dark clouds, which is not observed. Other bottom-up routes either shift the problem to the formation of intermediate ions, for which the formation through ion-molecule reactions starting from the protonation of C$_2$H$_2$ is currently under debate \citep{Kocheril2025}; rely on precursors that are observed but in low abundance (e.g. 1,3-butadiene; \citealt{agundez2025}); or are inefficient under interstellar conditions due to the lack of stabilising third bodies (\citealt{mcguire2018}). Consistently, chemical models incorporating these bottom-up pathways systematically underpredict the observed abundances of benzene and benzonitrile (\citealt{mcguire2018,burkhardt2021}).
This suggests that there are still important missing bottom-up routes not yet included in the models or, alternatively, that benzene can be rather synthesized through top-down mechanisms.

The bottom-up scenario has been favored by \citet{cernicharo2023_maps}. These authors proposed that the similar spatial distribution of \benzo with carbon chains such as \ce{HC3N}, \ce{HC5N} and \ce{HC7N} implies that aromatic molecules can be formed from chemical reactions involving smaller species.

In the top-down scenario the formation of aromatic species would proceed through the destruction of carbonaceous dust grains (e.g. \citealt{jones1996,scott1997,merino2014}), and/or via the fragmentation of large PAHs formed in the hot circumstellar medium (CSM) around evolved stars (e.g. \citealt{zhao2019}). The former process may indeed be viable in environments affected by shocks capable of sputtering the dust grains \citep{Berne2015}, like the CMZ molecular clouds. However, this mechanism cannot be easily invoked in cold dark clouds where \benzo (and larger aromatics in the case of TMC-1) has been detected with abundances comparable to those observed in CMZ clouds. 

\begin{figure}
    \centering
    \includegraphics[width=\linewidth]{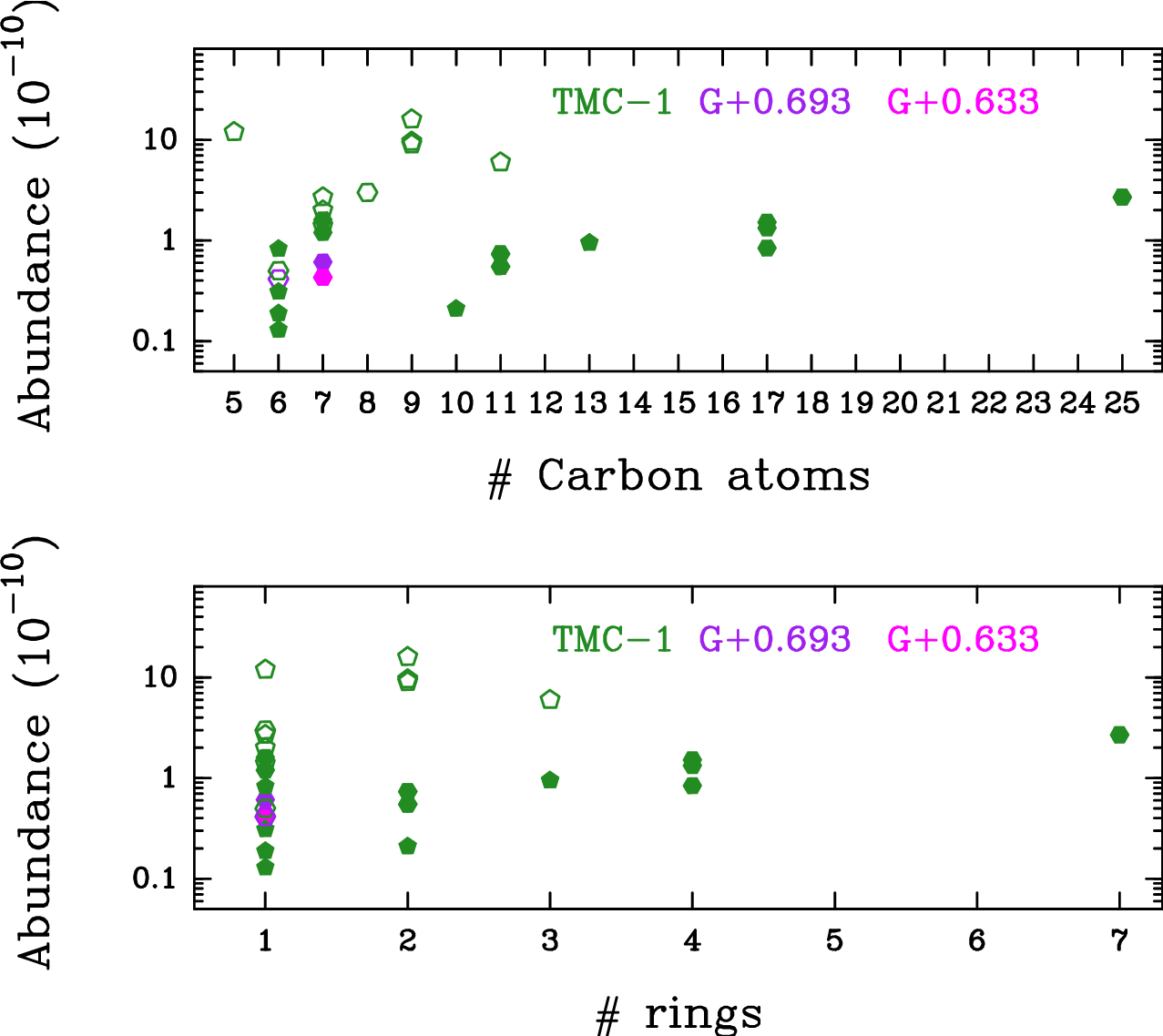}
    \caption{Molecular abundances of cyclic hydrocarbons detected in the ISM as a function of the number of carbon atoms (upper panel) and rings (lower panel). The colors indicate the different sources in which the species have been detected: green for TMC-1 and purple and magenta for G+0.693 and G+0.633. 
    Hexagons indicate aromatic compounds with only 6-carbon rings, while pentagons indicate compounds with (at least) one 5-carbon ring.
    The filled symbols denote CN-derivatives, which include the 6-carbon rings (\benzo, 1- and 2-c-\ce{C10H7CN}, 1- and 2- \ce{C16H9CN}, c-\ce{C24H11CN}, and c-\ce{C24H11CN}), and species with at least one 5-carbon ring (1- and 2-c-\ce{C5H5CN}, c-\ce{C9H7CN}, and 1- and 5-\ce{C12H7CN}). The open green symbols correspond to the cyclic hydrocarbons o-\ce{C6H4},  c-\ce{C5H6}, c-\ce{C5H4CCH2}, c-\ce{C9H8}, c-\ce{C6H5CCH}, c-\ce{C5H5CCH} and c-\ce{C11H8}.
    The open purple hexagon correspond to c-\ce{C6H6S}.
See references in the text.}
    \label{fig:PAHs} 
\end{figure}

Given the relatively large number of cyclic hydrocarbons detected in the ISM (21 so far) with increasing complexity, we can search for possible trends.
In Fig. \ref{fig:PAHs}, we show the molecular abundances of all interstellar cyclic hydrocarbons, as a function of the number of C atoms and C rings. We show the values of the abundances of \benzo reported towards the two CMZ clouds in this work, and those detected towards TMC-1 from the following works: \citet{burkhardt2021,cernicharo2021_benzyne,cernicharo2021_indene,cernicharo2021_ethynylcyclopentadiene,lee2021,mcguire2021,sita2022,cernicharo2022_fulvenallene,loru2023,cernicharo2024,Wenzel2024a,Wenzel2024b,wenzel2025,fuentetaja2026}.
For the six species reported both in the QUIJOTE and GOTHAM values (benzonitrile, 1- and 2-cyanocyclopentadiene, 1- and 2-cyanonapthalene, and indene), we compiled both values. The column densities derived by GOTHAM are usually larger by factors of 1.3$-$2.7, with the only exception of indene, which is a factor of 0.6 smaller. These discrepancies likely arise because the two teams used datasets covering different spectral ranges, and they apply different analysis. QUIJOTE assumed a single velocity component and that the emission fills the beam (which is supported by the maps presented in \citealt{cernicharo2023_maps}), as we have done for the CMZ clouds. GOTHAM performed the fit using 4 velocity components, leaving the source size as a free parameter in their Markov Chain Monte Carlo (MCMC) analysis, and thus they applied beam dilution. Most of the sizes derived (108$^{\prime\prime}-$413$^{\prime\prime}$) are larger that the GBT beams (22$^{\prime\prime}-$95$^{\prime\prime}$), so the average beam dilution factors are close to 1 (1$-$1.3). If beam dilutions had not been applied, the GOTHAM values would decrease by these values, being more consistent with those derived by QUIJOTE. Only for two molecules, which correspond to the larger species identified (the three isomers of cyanopyrene and cyanocoronene, with 4 and 7 rings respectively), the derived source size of 50$^{\prime\prime}$ is comparable/smaller than the GBT beam. For them, the beam filling factor is around $\sim$3, so the reported column density might be overestimated by this factor if the molecular emission is more extended. 

A remarkable aspect arising from Fig. \ref{fig:PAHs}, already noted by \citet{wenzel2025}, is that
the observed abundance of CN-derivatives (denoted by filled symbols) exhibit a flat (or even increasing) behavior with respect to molecular size, with multiple rings 
such as cyanopyrene (a 4-ring cyano-PAH) and even cyanocoronene (a 7-ring cyano-PAH) are similarly abundant than the single-ring \benzo.
Even if the abundances of these large species were overestimated by a factor of $\sim$3, as discussed above, the behavior would still be nearly flat, with cyanocoronene being almost equally abundant ($\sim$10$^{-10}$) than \benzo in TMC-1.

This observational fact differs significantly from the trends observed for other families of non-cyclic interstellar species, which are thought to be formed via a bottom-up scenario, namely, the most complex species from simpler ones.
This occurs for alcohols, thiols, aldehydes, or isocyanates, in which an increase in chemical complexity (understood as adding carbon atoms)
implies a decrease in the abundances of a factor of a few (e.g. \citealt{Rodriguez-Almeida2021a,Rodriguez-Almeida2021b,Jimenez-Serra2022,sanz-novo2022a}; and references therein). Another example of this is depicted in Fig. \ref{fig:cyano} for the cyanoployynes \ce{HC3N}, \ce{HC5N} and \ce{HC7N} towards both TMC-1 and the two CMZ sources. The column densities of TMC-1 have been obtained from \citet{xue2025};
the values of G+0.693 are from \citet{Colzi2024}, Colzi et al. (in prep.) and this work (Appendix \ref{sec:appendix-hc7n}); and the values of G+0.633 are from San Andrés et al. (in prep.) and this work (Appendix \ref{sec:appendix-hc7n} and \ref{sec:appendix-hc5n_G0633}). In all three cases, the increase of C atoms in the linear chain implies a clear decrease of the abundances. 
However, the trend observed for the cyclic hydrocarbons (Fig. \ref{fig:PAHs}) is strikingly different, which makes it difficult to envision \benzo (or simpler rings) as a direct precursor of the largest PAHs. 

Instead, a top-down scenario in which PAHs are continuously replenished by fragmentation of larger carbonaceous solids might explain better the similar abundances observed between small and large aromatics across different environments \citep{Seok2014}. This route integrates PAH formation into the global life cycle of dust, rather than requiring separate chemical pathways for each aromatic species. If dust grains can survive the transition from circumstellar envelopes through diffuse clouds to dense molecular clouds, the aromatic material they contain may serve as the natal reservoir for PAHs of all sizes. To allow this, PAHs need to survive in very different environments, including harsh physical conditions. 
In fact, PAHs are present in strongly irradiated environments such as photodissociation regions (PDR), with the Orion Bar being a prime example \citep{Peeters2002,JoblinTielens2011,knight2022}. Moreover, the observed infrared PAH bands (3.3, 6.2, 7.7, 8.6, 11.2 $\upmu$m) are widespread throughout the ISM, and spatial changes in band strength in PDRs support processing rather than new synthesis (\citealt{Berne2015}). 
In this context, the detection of \benzo in the harsh environment of the CMZ (for the first time besides quiescent cold clouds)  
provides compelling further observational evidence for the resilience of aromatic species across a wide range of physical conditions, including the presence of radiation, shocks and elevated cosmic-ray ionization rates. 

Recent laboratory and theoretical studies further support this hypothesis. Although PAHs are subject to destruction by ion collisions, their carbonaceous backbone can be preserved through efficient radiative cooling following energetic processing \citep{Stockett2023,Bull2025,Subramani2026}. 
Moreover, theoretical calculations and experimental evidence have shown that larger PAHs are more resilient to ultraviolet irradiation, shocks and cosmic-ray impacts than smaller ones (e.g., \citealt{Allain1996,Micelotta2010,Micelotta2011,Zhen2016}), which might also help to explain, at least partially, the trend shown in Fig. \ref{fig:PAHs}.
Concerning potential depletion onto grains, which would remove the PAHs from the gas phase, \citet{Dartois2022} investigated the sputtering yields of solid-phase perylene and coronene under energetic ion bombardment, analogous to cosmic-ray processing in dust-grain ice mantles. They demonstrated that cosmic-ray-induced sputtering can maintain a non-negligible gas-phase population of large PAHs, with predicted fractional abundances of coronene exceeding $10^{-10}$, which is consistent with observations in TMC-1 (\citealt{wenzel2025}; see Fig. \ref{fig:PAHs}). Besides, \citet{Piacentino2025} recently suggested the survival of aromatics in protoplanetary disks based on UV photodestruction experiments of small aromatics in both undiluted ices and astrophysically relevant ice matrices (H$_2$O, CO, CO$_2$).

All these findings support a substantial chemical inheritance of PAHs produced in the ISM (\citealt{Wenzel2024a}), and the evidence for such interstellar thread is also supported by findings in our Solar System. Recent measurements in asteroid Ryugu by \citet{Zeichner2023} have detected 2-, 3- and 4-membered ring PAHs. For the 2- and 4-membered rings, the isotopic clumping that points to a cold, interstellar origin, while the 3-membered rings were proposed to be formed by synthesis or reprocessing within moderate- to high-temperature settings, such as circumstellar environments or the parent body. Whatever the case, these detections in Ryugu shows that PAHs can be incorporated to protosolar nebula and survive the parent body processing.

\subsection{The role of aromatics in the C budget in molecular clouds}
\label{sec:discussion-C-budget}

\begin{figure}
    \centering
    \includegraphics[width=\linewidth]{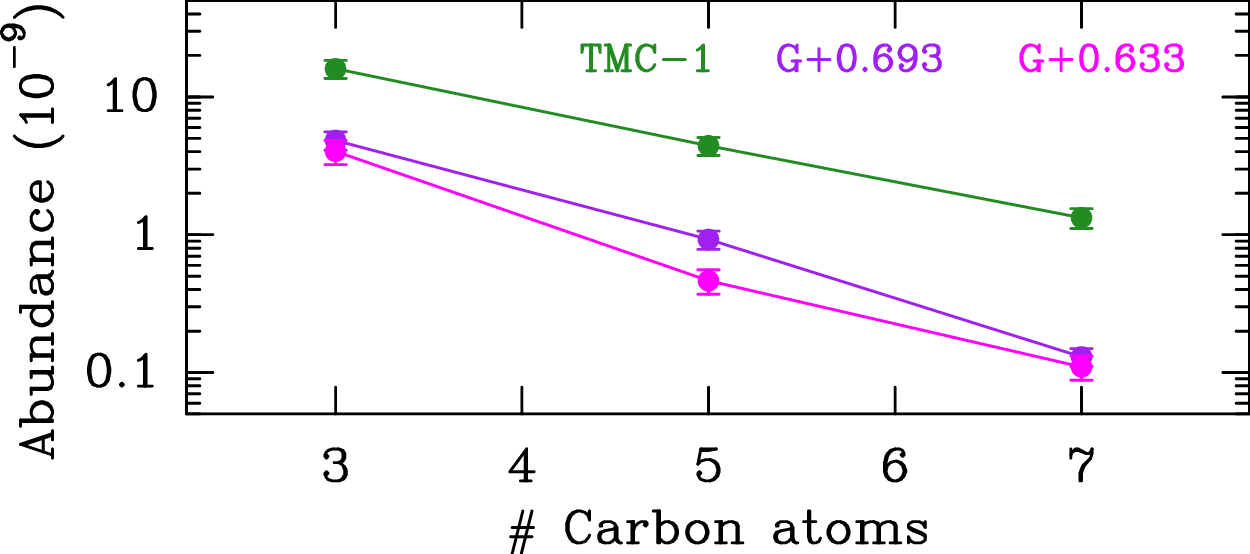}
    \caption{Molecular abundances of cyanopolyynes with 3,5 and 7 carbons (\ce{HC3N}, \ce{HC5N} and \ce{HC7N}). The column densities of TMC-1 have been obtained from \citet{xue2025};
    the values of G+0.693 are from \citet{Colzi2024}, Colzi et al. (in prep.) and this work (Appendix \ref{sec:appendix-hc7n}); the values of G+0.633 are from San Andrés et al. (in prep.) and this work (Appendix \ref{sec:appendix-hc7n} for \ce{HC7N} and Appendix \ref{sec:appendix-hc5n_G0633} for \ce{HC5N}). For the uncertainties we used the reported uncertainties of the column densities in each work, and assumed a 15$\%$ for the values of $N$(H$_2$).}
    \label{fig:cyano} 
\end{figure}

The increasing number of detections of aromatics and other cyclic hydrocarbons in different interstellar molecular clouds in recent years, and the fact that their abundances do not drop with increasing molecular size (and hence number of C atoms; see Fig. \ref{fig:PAHs}), unlike other C-carriers such as linear cyanopolyynes (see Fig. \ref{fig:cyano} ), might indicate that these species can contain a non-negligible amount of cosmic C in our Galaxy, as pointed out by \citet{Wenzel2024a}.
Moreover, given that the CMZ is often considered a local analogue of the molecular environments in other galaxies (e.g., \citealt{Langer2017}), these implications are also relevant for extragalactic contexts.

To have a rough estimate of this contribution, we can start with the simplest ring, benzene (c-\ce{C6H6}). To infer its abundance in the CMZ, we can use the derived abundance of its CN-derivative, \benzo, and assume a typical ratio between cyclic pure hydrocarbons and the CN derivatives (hereafter H/CN ratio).
For the pair c-\ce{C6H6}/\benzo, H/CN was predicted to be $\sim$10 according to the chemical models of \citet{Wenzel2024b}, assuming the rate constant derived by \citet{cooke2020}. 
Observationally, other H/CN ratios of 10-100 are found for several cyclic hydrocarbons in TMC-1 (\citealt{Wenzel2024a}). In G+0.693, the H/CN ratio can be estimated using the \ce{C4H3N} isomers. Both \ce{CH3CCCN} and \ce{CH2CCHCN} are expected to be formed from the combination of \ce{CH3CCH} and the CN radical (see discussion in \citealt{Rivilla2022_nitriles}). The ($N_{\ce{CH3CCCN}}$ + $N_{\ce{CH2CCHCN}}$) / $N_{\ce{CH3CCH}}$ ratio is 46, very similar to that found in TMC-1 (\citealt{marcelino2021}), and consistent with the 10-100 ratios found for cyclic hydrocarbons in TMC-1.

Assuming thus an H/CN ratio of 10–100, the expected abundance of benzene in the CMZ would be in the range (0.5–5)$\times$10$^{-9}$. Considering that benzene contains six carbon atoms, this corresponds to a contribution for the total C budget of 0.0006$\%-$0.006$\%$. The upper range of this estimate is  similar to the contributions of the two most abundant cyclic hydrocarbons detected to date towards TMC-1 (\citealt{cernicharo2021_indene}), indene (c-\ce{C9H8}) and cyclopentadiene (c-\ce{C5H6}), which are $\sim$0.003$\%$ and $\sim$0.001$\%$, respectively.
Although these individual contributions might be negligible, the detection of species with multiple rings, and then much larger number of C atoms (up to 25 so far), and the fact that their molecular abundances do not decrease compared to simpler species (Fig. \ref{fig:PAHs}), imply that the total contribution of the C budget of these species might be significant.
Indeed, after applying the H/CN ratio to the CN-derivatives detected towards TMC-1 (filled green symbols), the predicted abundances of their corresponding cyclic hydrocarbons would equal, or even be larger, than those of the most abundant cyclic hydrocarbons detected so far, of the order of $\sim$10$^{-9}$ (Fig. \ref{fig:PAHs}).
For instance, the derived abundance of the CN-derivative of the 7-ring PAH coronene in TMC-1 (\citealt{wenzel2025}), which contains 25 C atoms, provides a contribution of $\sim$0.013$\%$-0.13$\%$.
If larger PAHs are present at similar abundances $-$ which is a plausible scenario based on current evidence $-$, and considering the rapidly increasing number of C atoms in these species, the total contribution can be indeed relevant.
Moreover, we note that these estimates are likely lower limit values because of several reasons: i) they only consider only the gas-phase abundances, and it is expected that a fraction of PAHs is also locked in the surface of dust grains in molecular clouds; and ii) PAH derivatives (beyond thise with $-$CN) can also be present in the ISM, as demonstrated recently by the detection towards G+0.693 of the sulfur-bearing cycle c-C$_6$H$_6$S (\citealt{Araki2026}). Considering all this, it is not unreasonable that the total contribution of PAHs (and related cyclic species) to the C budget in molecular clouds could match the estimates from infrared dust models in the diffuse interstellar medium, which indicate that it can reach approximately 10$\%$–20$\%$ (e.g., \citealt{Li&Draine2001}; \citealt{Draine&Li2007}; \citealt{tielens2008}).

\section{Conclusions}\label{sec:conclusions}

The detection of \benzo in the CMZ molecular clouds G+0.693-0.027 and G+0.633-0.0604 demonstrates that this simple aromatic molecule can survive in the gas phase under the extreme physical conditions of the Central Molecular Zone of our Galaxy, characterized by the action of interstellar shocks and enhanced cosmic-ray ionization rates. Together with previous detections in cold dark clouds, these results firmly establish \benzo as a widespread constituent of the ISM, present across a broad range of Galactic environments.

We have found that the ratio between the unsaturated carbon chain \ce{HC7N} and \benzo is sensitive to  environmental conditions, being higher in colder sources (4.5$-$30) than in CMZ clouds ($\sim$2.3). This suggests that these two C-rich molecules are not directly chemically linked, and a preference of aromatic chemistry in the CMZ compared to linear C-chains.

Several bottom-up mechanisms proposed for the synthesis of benzene struggle to reproduce the observed abundances or to explain their weak dependence on physical conditions. Chemical models including these pathways systematically underpredict the abundances of benzene and \benzo, indicating that key processes are still missing or that additional formation channels must be considered. In contrast, a top-down scenario in which aromatic species are produced and replenished through the fragmentation and processing of larger carbonaceous solids or PAHs offers a coherent framework to explain the observations. The comparable abundances of single-ring aromatics and much larger PAHs detected in cold clouds, together with the survival of PAHs in strongly irradiated and shocked environments, support the idea that aromatic material is chemically resilient and continuously recycled throughout the ISM. In this picture, \benzo may act as a tracer of aromatic fragments originating from the processing of larger PAHs, linking small aromatic molecules to the global life cycle of interstellar carbonaceous material. Moreover, these findings highlight the need for astrochemical models to explicitly incorporate top-down processes.

The presence of \benzo in the CMZ, together with very recent observational and experimental results on related cycles \citep{Araki2026}, opens the way to the detection of more complex aromatic molecules in the Galactic Center, such as those recently identified towards the dark molecular cloud TMC-1 \citep{Wenzel2024a,Wenzel2024b,wenzel2025}, or even more functionalized cycles \citep{Piacentino2026}. Finally, although the direct contribution of \benzo itself to the carbon budget of the ISM is modest, if larger PAHs are present at similar levels, their cumulative contribution could approach a significant fraction of the total interstellar carbon. 
This turns aromatic chemistry into an integral component of the Galactic carbon cycle, which should not be ruled out in astrochemical models.

\begin{acknowledgements}
We are grateful to the Yebes 40m and IRAM 30m 
telescope staff for their help during the different observing runs. The 40m radio telescope at Yebes Observatory is operated by the Spanish Geographic Institute (IGN, Ministerio de Transportes, Movilidad y Agenda Urbana). IRAM is supported by INSU/CNRS (France), MPG (Germany) and IGN (Spain). 
V.M.R., I.J.S, and L.C. acknowledge support from the grant PID2022-136814NB-I00 by the Spanish Ministry of Science, Innovation and Universities/State Agency of Research MICIU/AEI/10.13039/501100011033 and by ERDF, UE.
V.M.R. also aknowledges the grant RYC2020-029387-I funded by MICIU/AEI/10.13039/501100011033 and by "ESF, Investing in your future", and from the Consejo Superior de Investigaciones Cient{\'i}ficas (CSIC) and the Centro de Astrobiolog{\'i}a (CAB) through the project 20225AT015 (Proyectos intramurales especiales del CSIC). V.M.R., D.S.A and M.S.-N. acknowledge support from the grant CNS2023-144464 funded by MICIU/AEI/10.13039/501100011033 and by “European Union NextGenerationEU/PRTR”. D.S.A. also expresses his gratitude for the funds received from the Comunidad de Madrid through the Grant PIPF-2022/TEC-25475 and the financial support provided by the Consejo Superior de Investigaciones Cient{\'i}ficas (CSIC) and the Centro de Astrobiolog{\'i}a (CAB) through the project 20225AT015 (Proyectos intramurales especiales del CSIC). M.S.-N. acknowledges funding from the Alexander von Humboldt foundation under a Humboldt Research Fellowship.
I.J-.S acknowledges funding from grant PID2022-136814NB-I00 funded by the Spanish Ministry of Science, Innovation and Universities/State Agency of Research MICIU/AEI/ 10.13039/501100011033 and by “ERDF/EU”. I.J.-S. and A.M. acknowledge support from the ERC CoG grant OPENS (project number GA 101125858) funded by the European Union.
B. T. thanks the Spanish MICIU for funding support from grants PID2022-137980NB-I00 and PID2023-147545NB-I00.
The project that gave rise to these results received the support of a fellowship from the ”la Caixa” Foundation (ID 100010434). The fellowship code is LCF/BQ/PR25/12110012.
J.L. acknowledges support from KU Leuven through Project No. C14/22/082.

\end{acknowledgements}

\bibliographystyle{aa}
\bibliography{biblio}

%
%

\begin{appendix}

\section{Analysis of HC$_7$N}
\label{sec:appendix-hc7n}

\begin{figure*}
    \centering
    \includegraphics[width=\textwidth]{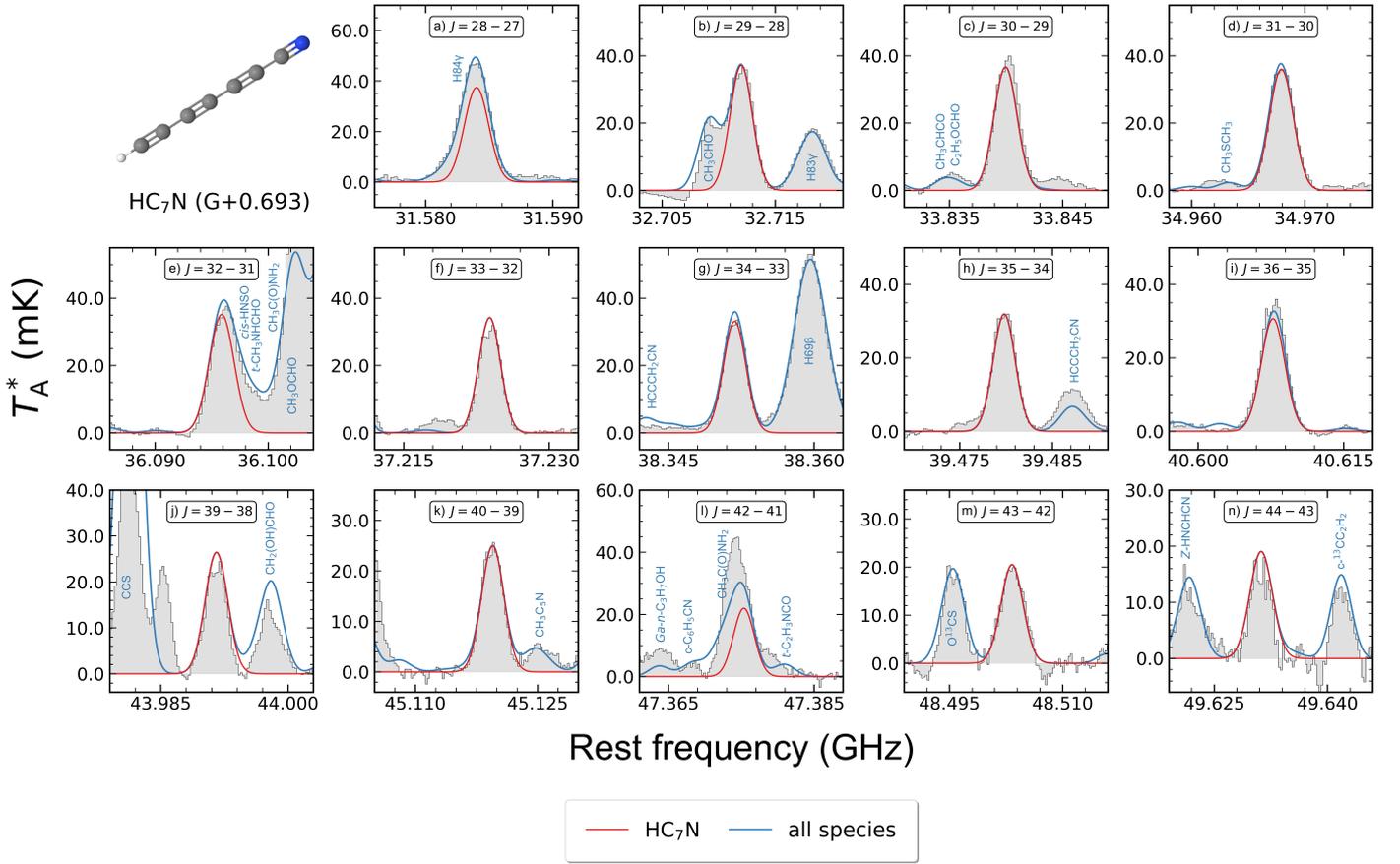}
    \caption{Unblended or slightly blended \ce{HC7N} transitions observed towards G+0.693. The black histogram and grey-shaded areas delineate the observed spectrum, while the red and blue solid lines outline the best LTE model obtained for \ce{HC7N} and the contribution from the rest of species already identified towards this cloud (with their corresponding labels), respectively. Panel labels refer to the specific rotational transitions being plotted, whose spectroscopic details are given in Table~\ref{tab:HC7N_transitions}.} 
    \label{fig:G+0.693_HC7N_low_J} 
\end{figure*}

\begin{figure*}
    \centering
    \includegraphics[width=\textwidth]{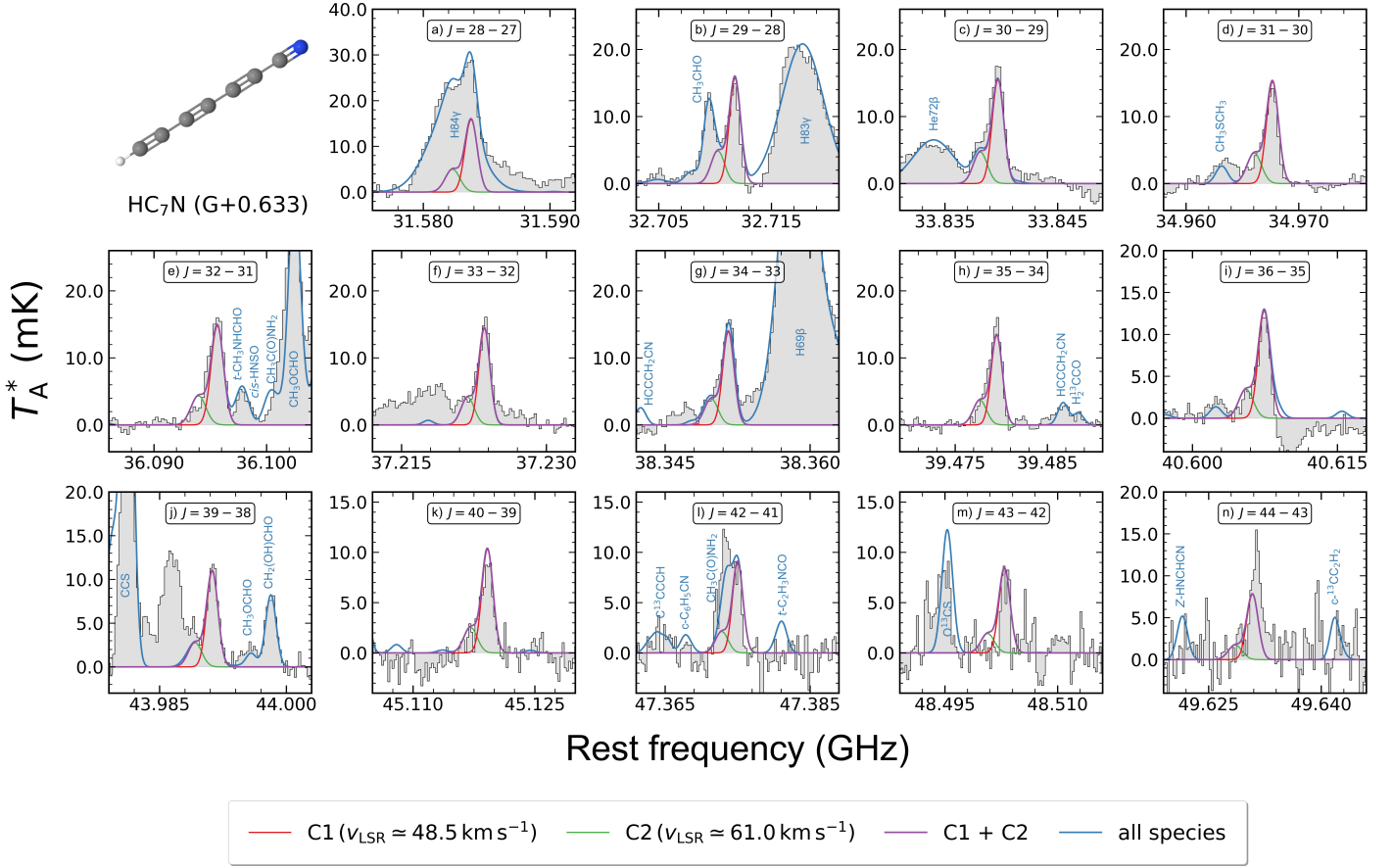}
    \caption{Same as Fig.~\ref{fig:G+0.693_HC7N_low_J} but for G+0.633, where the same set of rotational transitions has been identified and fitted. In this case, the red and green solid lines outline the best LTE model obtained for the two major velocity components already characterised in the cloud (C1 and C2, respectively; San Andr{\'e}s et al., in prep.), together with their combined profile in purple.} 
    \label{fig:G+0.633_HC7N_low_J} 
\end{figure*}

We have performed the analysis of \ce{HC7N} towards G+0.693 and G+0.633 by using the rotational spectroscopic data measured by \citet{Giesen2020}, \citet{Bizzocchi&DegliEsposti2004} and \citet{Kirby1980} (entry 99501 from the CDMS catalog). In both sources, we have detected the whole progression ranging from $J_\text{up} = 28$ up to $J_\text{up} = 44$, with the exception of the $J_\text{up} = 37, 38$ transitions that appear heavily blended with yet unknown species and the $J_\text{up} = 41$ line that suffers heavy blending from \ce{^{13}CS}. More energetic \ce{HC7N} lines (with $J_\text{up} \geq 64$), present in the spectral surveys of both clouds, are not properly identified in any of them, and consequently excluded in this species analysis. Figs.~\ref{fig:G+0.693_HC7N_low_J} and ~\ref{fig:G+0.633_HC7N_low_J} show the LTE analysis performed for this molecule towards G+0.693 and G+0.633, respectively, using the SLIM-AUTOFIT tool of MADCUBA. While in G+0.693 only one velocity component has been fitted, in G+0.633 the profiles of \ce{HC7N} traced two distinct velocity components as already characterised for many other species in this cloud (San Andr{\'e}s et al., in prep.). The results of the fittings in both clouds are summarised in Table~\ref{tab:HC7N_fitting}. We obtained the following in G+0.693: $N = (1.72 \pm 0.03)\times10^{13}\, \text{cm}^{-2}$, $T_\text{ex} = 19.8 \pm 0.4$\K, $v_\text{LSR} = 66.1 \pm 0.1$\kms and $\text{FWHM} = 21.3 \pm 0.3$\kms, which are consistent with those reported in a previous work (e.g., $N = (1.5 \pm 0.3)\times10^{13}\, \text{cm}^{-2}$; \citealt{Zeng2018}). Meanwhile, we derived in G+0.633: $N = (6.3 \pm 0.4)\times10^{12}\, \text{cm}^{-2}$, $T_\text{ex} = 19.3 \pm 1.1$\K, $v_\text{LSR} = 48.8 \pm 0.3$\kms and $\text{FWHM} = 10.8 \pm 0.5$\kms for the main component (C1), and $N = (2.6 \pm 0.6)\times10^{12}\, \text{cm}^{-2}$, $T_\text{ex} = 16 \pm 3$\K and $v_\text{LSR} = 63 \pm 1$\kms for the secondary one (C2), fixing the $\text{FWHM}$ to 13\kms, as typically found for this component (San Andr{\'e}s et al., in prep.), to allow for convergence. The derived abundances with respect to \ce{H2} for \ce{HC7N} are (1.3$\pm$0.2)$\times$10$^{-10}$ for G+0.693, and (1.1$\pm$0.2)$\times$10$^{-10}$ and (3.6$\pm$1.3)$\times$10$^{-11}$ for the C1 and C2 components identified in G+0.633, respectively. For G+0.693 we used $N_{\ce{H2}} = 1.35\times10^{23}$ cm$^{-2}$ \citep{Martin2008} considering a 15\% of its value as uncertainty; for G+0.633 we employed $N_{\ce{H2}} = (6.0 \pm 1.2)\times10^{22}$ cm$^{-2}$ for C1 and $N_{\ce{H2}} = (7 \pm 2)\times10^{22}$ cm$^{-2}$ for C2 (San Andrés et al., in prep.).

\begin{table}
\tabcolsep 0.5pt
\centering
\caption{Derived physical parameters of \ce{HC7N} towards G+0.693 and G+0.633. The parameters without uncertainties were fixed in the fit.}
\begin{tabular}{ c c c c c c }
\hline
Cloud$^a$  & $N$ &  $T_{\rm ex}$ & $v_{\rm LSR}$ & FWHM  & $\chi$$^b$    \\
  & ($\times$10$^{12}$ cm$^{-2}$) & (K) & (km s$^{-1}$) & (km s$^{-1}$) & ($\times$10$^{-10}$)      \\
\hline
G+0.693 & 17.2$\pm$0.3 & 19.8$\pm$0.4 & 66.1$\pm$0.1 & 21.3$\pm$0.3 & 1.3$\pm$0.2 \\
G+0.633-C1 & 6.3$\pm$0.4 & 19.3$\pm$1.1 & 48.8$\pm$0.3 & 10.8$\pm$0.5 & 1.1$\pm$0.2 \\
G+0.633-C2 & 2.6$\pm$0.6 & 16$\pm$3 & 63$\pm$1 & 13.0 & 0.36$\pm$0.13 \\
\hline
\\[-7pt]
\multicolumn{6}{p{.45\textwidth}}{$^a$ For G+0.633, additional labels indicating the velocity component (C1 or C2) are included. $^b$ For G+0.693, we employed $N_{\rm H_2}$ = 1.35$\times$10$^{23}$ cm$^{-2}$, from \citet{Martin2008}, assuming an uncertainty of 15\% of its value; For G+0.633 we used $N_{\rm H_2}$ = (0.60$\pm$0.12)$\times$10$^{23}$ cm$^{-2}$ for C1 and $N_{\rm H_2}$ = (0.7$\pm$0.2)$\times$10$^{23}$ cm$^{-2}$ for C2 (San Andrés et al., in prep.). 
}
\end{tabular}
\label{tab:HC7N_fitting}
\end{table}

\begin{table*}
\centering
\caption{List of transitions of \ce{HC7N} identified towards both the G+0.693 (Fig.~\ref{fig:G+0.693_HC7N_low_J}) and G+0.633 (Fig.~\ref{fig:G+0.633_HC7N_low_J}) molecular clouds. For each transition, we provide its rest frequency (with its corresponding uncertainty associated to the last digits indicated into brackets), quantum numbers, base 10 logarithm of the integrated intensity at 300 K (log $I$) and the values of the lower and upper energy levels involved in of each transitions ($E_{\rm l}$ and $E_{\rm u}$), in cm$^{-1}$ and K, respectively. The last column provides the blending information of the identified lines.}
\begin{tabular}{ l c c c c c }
\hline
Frequency & Transition & log $I$ & $E_{\rm l}$ & $E_{\rm u}$ & Blending$^a$ \\
(GHz) &   & (nm$^2$ MHz) & (cm$^{-1}$) & (K) &  \\
\hline
31.58370(1) & $28-27$ & -3.4399 & 14.2 & 22.0 & H84$\gamma$ \\
32.71168(1) & $29-28$ & -3.3964 & 15.3 & 23.5 & Unblended \\
33.83962(1) & $30-29$ & -3.3546 & 16.4 & 25.2 & Unblended \\
34.96759(1) & $31-30$ & -3.3142 & 17.5 & 26.9 & Unblended \\
36.09553(1) & $32-31$ & -3.2753 & 18.7 & 28.6 & Unblended \\
37.22349(1) & $33-32$ & -3.2378 & 19.9 & 30.4 & Unblended \\
38.35145(1) & $34-33$ & -3.2015 & 21.1 & 32.2 & Unblended \\
39.47941(1) & $35-34$ & -3.1665 & 22.4 & 34.1 & Unblended \\
40.6073267(5) & $36-35$ & -3.1325 & 23.7 & 36.1 & Unblended \\
43.9911288(5) & $39-38$ & -3.0371 & 27.9 & 42.2 & Unblended \\
45.1190554(5) & $40-39$ & -3.0072 & 29.3 & 44.4 & Unblended \\
47.3748968(6) & $42-41$ & -2.9500 & 32.4 & 48.9 & \ce{CH3C(O)NH2}, U-line \\
48.5028115(6) & $43-42$ & -2.9227 & 34.0 & 51.2 & Unblended \\
49.6307220(6) & $44-43$ & -2.8962 & 35.6 & 53.6 & U-line \\
\hline 
\end{tabular}
\label{tab:HC7N_transitions}
{\\ (a) Given that line blending is exactly the same for both G+0.693 and G+0.633, no separation between both clouds is made. We note that in the case of G+0.633, blending information is only considered for the profile related to the main C1 velocity component (red line). When applicable, ``U'' refers to blending with emission from an unknown (not yet identified) species.}
\end{table*}

\section{Analysis of \ce{HC5N} towards G+0.633}
\label{sec:appendix-hc5n_G0633}

\begin{figure*}
    \centering
    \includegraphics[width=\textwidth]{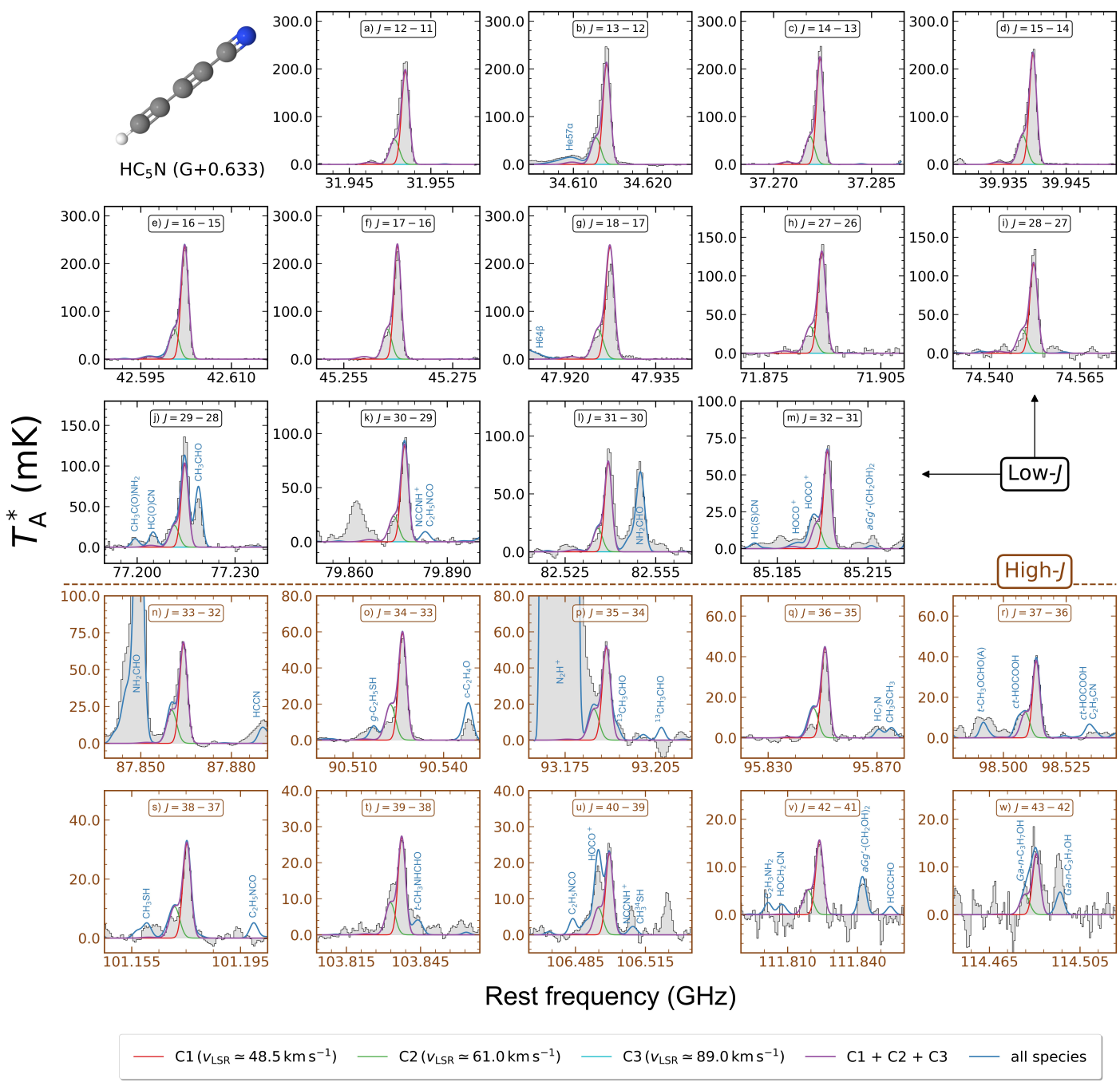}
    \caption{Lines of \ce{HC5N} observed observed towards the G+0.633 molecular cloud, separated into the ``low-$J$'' ($E_{\text{up}} \leq 70$\K, upper panels in black) and ``high-$J$'' ($E_{\text{up}} > 70$\K, bottom panels in brown) subgroups, for which the fitting was performed independently as done for G+0.693 (Colzi et al., in prep.). The black histogram and grey-shaded areas delineate the observed spectrum. The red, green and cyan solid lines outline the LTE fitting for the C1, C2 and C3 components identified in this cloud, respectively. In blue, it is shown the contribution from the rest of species already identified towards this cloud thus far (with their corresponding labels). Panel labels refer to the specific rotational transitions being plotted, whose spectroscopic details are given in Table~\ref{tab:HC5N_transitions}.}
    \label{fig:HC5N_G0633}
\end{figure*}

In this appendix, we present the LTE analysis on \ce{HC5N} we performed towards G+0.633, using the rotational spectroscopic data provided by \citet{Alexander1976}, \citet{Winnewisser1982}, \citet{Bizzocchi2004} and \citet{Giesen2020}. We identified nearly two dozen lines, divided into two sets of transitions that span rotational levels $12 \leq J_{\text{up}} \leq 18$ and $27 \leq J_{\text{up}} \leq 43$, excluding the $J_{\text{up}} = 41$ line that is heavily blended with \ce{HC3N}. The rest of \ce{HC5N} transitions covered in the current survey of G+0.633 (with $J_{\text{up}} \geq 79$), are too faint to be detected, and therefore not considered for the analysis of this species. We show in Fig.~\ref{fig:HC5N_G0633} all \ce{HC5N} transitions reported in G+0.633, which were analysed following the same approach as adopted in G+0.693 (see Colzi et al., in prep.). Because of non-LTE effects, all these 23 transitions cannot be consistently reproduced attending to one LTE model. Thus, we performed their fitting by separating them into two groups: (i) $5 \leq E_{\text{up}} \leq 70$\K (the ``low-$J$'' group, corresponding to transitions with $J_{\text{up}}$ between 12 and 32), and (ii) $E_{\text{up}} > 70$\K (``high-$J$'', delineating lines with $J_{\text{up}}$ between 33 and 43). The results for each individual group are collected in Table~\ref{tab:HC5N_fitting}.

The profiles of the ``low-$J$'' lines outline all three velocity components identified in G+0.633 (C1, the main component, observed for both \benzo and \ce{HC7N}; C2, the secondary one, also traced by \ce{HC7N}; and C3, the faintest among the three; San Andrés et al., in prep.), although the contribution of C3 is only discernible for the lowest energy transitions (up to the $J_{\text{up}} = 18$ line). We performed the fitting of the three components simultaneously. For the C1 component, we allowed all physical parameters to vary free, deriving $N = (2.78 \pm 0.03)\times10^{13}\, \text{cm}^{-2}$, $T_\text{ex} = 18.5 \pm 0.2$\K, $v_\text{LSR} = 48.50 \pm 0.05$\kms and $\text{FWHM} = 10.20 \pm 0.13$\kms. As for the C2 and C3 components, we fixed the $T_{\text{ex}}$ (18.5\K for both of them, as retrieved for C1), $v_{\text{LSR}}$ (61\kms and 89\kms, respectively) and $\text{FWHM}$ (13\kms and 18\kms, respectively) to allow for convergence, obtaining $N = (9.4 \pm 0.3)\times10^{12}\, \text{cm}^{-2}$ for C2 and $N = (1.1 \pm 0.3)\times10^{12}\, \text{cm}^{-2}$ for C3. The derived column densities translate into abundances relative to \ce{H2} of $(4.6 \pm 0.9) \times 10^{-10}$ for C1, $(1.3 \pm 0.4) \times 10^{-10}$ for C2 and $(3.9 \pm 1.5) \times 10^{-11}$ for C3. The following $N_{\rm H_2}$ values were used: (6.0$\pm$1.2)$\times$10$^{22}$ cm$^{-2}$ for C1, (7$\pm$2)$\times$10$^{22}$ cm$^{-2}$ for C2 and (2.8$\pm$0.7)$\times$10$^{22}$ cm$^{-2}$ for C3 (San Andrés et al., in prep.).

As for the ``high-$J$'' lines, these are only detected in C1 and C2, and we performed their fitting similarly to the ``low-$J$'' transitions. For the C1 component, all four parameters were left free, yielding $N = (2.6 \pm 0.2)\times10^{13}\, \text{cm}^{-2}$, $T_\text{ex} = 22.1 \pm 0.8$\K, $v_\text{LSR} = 47.7 \pm 1.3$\kms and $\text{FWHM} = 10.3 \pm 0.3$\kms as the best estimates. For the C2 component, the $T_{\text{ex}}$, $v_{\text{LSR}}$ and $\text{FWHM}$ were fixed to 22.1\K, 61\kms and 13\kms, respectively, obtaining $N = (9 \pm 5)\times10^{12}\, \text{cm}^{-2}$. The resulting abundances with respect to \ce{H2} are $(4.3 \pm 0.9) \times 10^{-10}$ for C1 and $(1.3 \pm 0.8) \times 10^{-10}$ for C2.

\begin{table}
\tabcolsep 0.5pt
\centering
\caption{LTE analysis results of \ce{HC5N} towards G+0.633. The parameters without uncertainties were fixed in the fit.}
\begin{tabular}{ c c c c c c }
\hline
Comp.  & $N$ &  $T_{\rm ex}$ & $v_{\rm LSR}$ & FWHM  & $\chi$$^a$    \\
  & ($\times$10$^{13}$ cm$^{-2}$) & (K) & (km s$^{-1}$) & (km s$^{-1}$) & ($\times$10$^{-10}$)      \\
\hline
\multicolumn{6}{c}{Low-$J$ lines ($E_{\text{up}} \leq 70$\K)} \\
\hline
C1 & 2.78$\pm$0.03 & 18.5$\pm$0.2 & 48.50$\pm$0.05 & 10.20$\pm$0.13 & 4.6$\pm$0.9 \\
C2 & 0.94$\pm$0.03 & 18.5 & 61.0 & 13.0 & 1.3$\pm$0.4 \\
C3 & 0.11$\pm$0.03 & 18.5 & 89.0 & 18.0 & 0.39$\pm$0.15 \\
\hline
\multicolumn{6}{c}{High-$J$ lines ($E_{\text{up}} > 70$\K)} \\
\hline
C1 & 2.6$\pm$0.2 & 22.1$\pm$0.8 & 47.7$\pm$1.3 & 10.3$\pm$0.3 & 4.3$\pm$0.9 \\
C2 & 0.9$\pm$0.5 & 22.1 & 61.0 & 13.0 & 1.3$\pm$0.8 \\
\hline
\\[-7pt]
\multicolumn{6}{p{.45\textwidth}}{$^a$ We used the following $N_{\rm H_2}$ values: (6.0$\pm$1.2)$\times$10$^{22}$ cm$^{-2}$ for C1, (7$\pm$2)$\times$10$^{22}$ cm$^{-2}$ for C2 and (2.8$\pm$0.7)$\times$10$^{22}$ cm$^{-2}$ for C3 (San Andrés et al., in prep.). 
}
\end{tabular}
\label{tab:HC5N_fitting}
\end{table}

\begin{table*}
\centering
\caption{List of transitions of \ce{HC5N} identified towards the G+0.633 molecular cloud and shown in Fig.~\ref{fig:HC5N_G0633}). For each transition, we provide its rest frequency (with its corresponding uncertainty associated to the last digits indicated into brackets), quantum numbers, base 10 logarithm of the integrated intensity at 300 K (log $I$) and the values of the lower and upper energy levels involved in of each transitions ($E_{\rm l}$ and $E_{\rm u}$), in cm$^{-1}$ and K, respectively. The last column provides the blending information of the identified lines, taking only the main C1 component into consideration (red line).}
\begin{tabular}{ l c c c c c }
\hline
Frequency & Transition & log $I$ & $E_{\rm l}$ & $E_{\rm u}$ & Blending$^a$ \\
(GHz) &   & (nm$^2$ MHz) & (cm$^{-1}$) & (K) &  \\
\hline
\multicolumn{6}{c}{Low-$J$ lines ($E_{\text{up}} \leq 70$\K)} \\
\hline
31.95177(1) & $12-11$ & -3.5006 & 5.9 & 10.0 & Unblended \\
34.61439(1) & $13-12$ & -3.3986 & 6.9 & 11.6 & Unblended \\
37.27699(1) & $14-13$ & -3.3045 & 8.1 & 13.4 & Unblended \\
39.93959(1) & $15-14$ & -3.2173 & 9.3 & 15.3 & Unblended \\
42.6021529(1) & $16-15$ & -3.1361 & 10.7 & 17.4 & Unblended \\
45.2647199(1) & $17-16$ & -3.0602 & 12.1 & 19.6 & Unblended \\
47.9272746(1) & $18-17$ & -2.9890 & 13.6 & 21.9 & Unblended \\
71.8895950(2) & $27-26$ & -2.4982 & 31.2 & 48.3 & Unblended \\
74.5519871(2) & $28-27$ & -2.4559 & 33.6 & 51.9 & Unblended \\
77.2143590(2) & $29-28$ & -2.4154 & 36.1 & 55.6 & \ce{CH3CHO} \\
79.876710(7) & $30-29$ & -2.3767 & 38.6 & 59.4 & Unblended \\
82.5390393(2) & $31-30$ & -2.3396 & 41.3 & 63.4 & Unblended \\
85.201340(7) & $32-31$ & -2.3041 & 44.1 & 67.5 & Unblended \\
\hline
\multicolumn{6}{c}{High-$J$ lines ($E_{\text{up}} > 70$\K)} \\
\hline 
87.8636300(2) & $33-32$ & -2.2700 & 46.9 & 71.7 & Unblended \\
90.5258899(3) & $34-33$ & -2.2373 & 49.8 & 76.0 & Unblended \\
93.188123(7) & $35-34$ & -2.2059 & 52.8 & 80.5 & U-line \\
95.8503354(3) & $36-35$ & -2.1758 & 56.0 & 85.1 & Unblended \\
98.512524(7) & $37-36$ & -2.1469 & 59.2 & 89.8 & Unblended \\
101.1746768(3) & $38-37$ & -2.1190 & 62.4 & 94.7 & Unblended \\
103.836817(7) & $39-38$ & -2.0923 & 65.8 & 99.7 & U-line \\
106.498910(7) & $40-39$ & -2.0667 & 69.3 & 104.8 & \ce{HOCO+} \\
111.823025(3) & $42-41$ & -2.0183 & 76.5 & 115.4 & Unblended \\
114.4850330(7) & $43-42$ & -1.9955 & 80.2 & 120.9 & $Ga$-$n$-\ce{C3H7OH} \\
\hline
\end{tabular}
\label{tab:HC5N_transitions}
{\\ (a) ``U'' refers to blending with emission from an unknown (not yet identified) species.}
\end{table*}

\section{Collisional rates of \benzo with He up to J=19 and 100 K}
\label{sec:appendix-rates}

The rotational (de-)excitation of interstellar benzonitrile (c-C$_6$H$_5$CN) in collisions with helium atoms was recently investigated by \citet{khalifa2024} for rotational levels up to $J=9$ and for kinetic temperatures up to 40 K on an accurate potential energy surface (PES).
The interaction potential  is described using three Jacobi coordinates $(R,\theta,\phi)$, where $R$ is the intermolecular separation, and $\theta$ and $\phi$ are the angular coordinates defining the orientation of the He atom relative to the molecular frame. The \benzo molecule was treated as a rigid asymmetric top. A total of 15,000 \emph{ab initio} points were computed over 60 values of $R$ from 2 to 50~bohr,  25 values of $\theta$ (0--180$^\circ$), and 10 values of $\phi$ (0--90$^\circ$). This fine grid was necessary due to the strong anisotropy of the potential. 
All PES points were computed using the explicitly correlated coupled-cluster method CCSD(T)-F12a with the aug-cc-pVTZ basis set, as implemented in the, including a counterpoise correction to account for the basis set superposition error.
Because CCSD(T)-F12a is not perfectly size-consistent, the small residual long-range interaction energy was subtracted from all points to enforce correct asymptotic behavior. 
The PES exhibits a strongly anisotropic structure with a global minimum of $D_e = -97.2$~cm$^{-1}$ at $R = 3.1$~\AA, $\theta = 78^\circ$, $\phi = 90^\circ$. 
For use in scattering calculations, the PES was expanded in spherical harmonics:
\[
V(R,\theta,\phi) = \sum_{l=0}^{l_{\mathrm{max}}} \sum_{m=-l}^{l}
V_{lm}(R) \, Y_{l}^{m}(\theta,\phi),
\]
restricted by C$_{2v}$ symmetry to even-$m$ components. The final representation includes 110 expansion terms using $l_{\max}=19$, with an accuracy better than 1~cm$^{-1}$ for $R \ge 5$~bohr. 

State-to-state rotational excitation cross sections were computed using the quantum time-independent \emph{coupled states} (CS) approximation, implemented in \textsc{Molscat}. Benchmark comparisons showed that CS reproduces close-coupling (CC) results with typical errors below 15\%, while drastically reducing the computational time.
Separate calculations were carried out for \textit{ortho}- and \textit{para}-benzonitrile due to nuclear-spin symmetry. 

In the present work the calculations were extended to higher collision energies (up to 500 cm$^{-1}$) and higher rotational levels. Specifically, the rotational basis of \textit{para}-C$_6$H$_5$CN (resp. \textit{ortho}-C$_6$H$_5$CN) includes all levels up to $J_{\max}=20$ (resp. 19) for energies up to 100 cm$^{-1}$, $J_{\max}=21$ (resp. 20) for energies between 100 and 200 cm$^{-1}$, $J_{\max}=21$ (resp. 20) for energies between 100 and 200 cm$^{-1}$, $J_{\max}=22$ (resp. 21) for energies between 200 and 320 cm$^{-1}$, and $J_{\max}=23$ (resp. 22) for energies above 320 cm$^{-1}$. 
The calculations for energies lower than 200 cm$^{-1}$ are performed on the same grid of energies as reported in \citet{khalifa2024}, while at higher energies cross sections were computed every 10 cm$^{-1}$. 

State-to-state rate coefficients up to kinetic temperatures of 100 K were computed by thermally averaging the cross sections over a Maxwell--Boltzmann distribution.
The results show that the largest rate coefficients are associated with transitions with $\Delta K_a = 0$ both for \textit{ortho}- and \textit{para}-benzonitrile.

\end{appendix}

\end{document}